\documentclass[a4paper,11pt]{article}
\usepackage{jinstpub}
\usepackage{lineno}

\usepackage{amssymb}
\usepackage{lipsum}
\usepackage{url}
\usepackage{lineno}
\usepackage{soul}
\usepackage[dvipsnames]{xcolor}
\definecolor{darkred}{rgb}{0.5,0,0}
\definecolor{darkblue}{rgb}{0,0,0.5}
\definecolor{firebrick}{rgb}{0.75,0.125,0.125}
\definecolor{darkgreen}{rgb}{0,0.5,0}
\hypersetup{urlcolor=darkblue,
	    citecolor=darkgreen,
	    linkcolor=firebrick}

\begin{document}

\title{Towards the Giant Radio Array for Neutrino Detection (GRAND): the GRANDProto300 and GRAND@Auger prototypes}

\collaboration{\includegraphics[height=17mm]{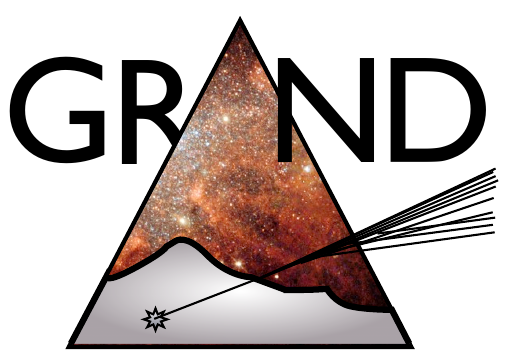}\\[6pt]
 GRAND Collaboration}

\author[a]{J.~Álvarez-Muñiz,}
\author[b,c]{R.~Alves Batista,}
\author[d]{A.~Benoit-Lévy,}
\author[e,f]{T.~Bister,}
\author[g]{M.~Bohacova,}
\author[h]{M.~Bustamante,}
\author[i]{W.~Carvalho,}
\author[j,k]{Y.~Chen,}
\author[l]{L.~Cheng,}
\author[m]{S.~Chiche,}
% ...

\affiliation[a]{Departamento de Física de Particulas \& Instituto Galego de Física de Altas Enerxías, Universidad de Santiago de Compostela, 15782 Santiago de Compostela, Spain}
\affiliation[b]{Institut d'Astrophysique de Paris, CNRS, Sorbonne Université, 98 bis bd Arago 75014, Paris, France}
\affiliation[c]{Sorbonne Université, Université Paris Diderot, Sorbonne Paris Cité, CNRS, Laboratoire de Physique Nucléaire et de Hautes Energies (LPNHE), 4 Place Jussieu, F-75252, Paris Cedex 5, France}
\affiliation[d]{Université Paris-Saclay, CEA, F-91120 Palaiseau, France}
\affiliation[e]{Institute for Mathematics, Astrophysics and Particle Physics, Radboud Universiteit, Nijmegen, the Netherlands}
\affiliation[f]{Nikhef, National Institute for Subatomic Physics, Amsterdam, the Netherlands}
\affiliation[g]{Institute of Physics of the Czech Academy of Sciences, Na Slovance 1999/2, 182 00 Prague 8, Czechia}
\affiliation[h]{Niels Bohr International Academy, Niels Bohr Institute, University of Copenhagen, 2100 Copenhagen, Denmark}
\affiliation[i]{Faculty of Physics, University of Warsaw, Pasteura 5, 02-093 Warsaw, Poland}
\affiliation[j]{Key Laboratory of Dark Matter and Space Astronomy, Purple Mountain Observatory, Chinese Academy of Sciences, 210023 Nanjing, Jiangsu, China}
\affiliation[k]{School of Astronomy and Space Science, University of Science and Technology of China, 230026 Hefei Anhui, China}
\affiliation[l]{National Astronomical Observatories, Chinese Academy of Sciences, Beijing 100101, China}
\affiliation[m]{Inter-University Institute For High Energies (IIHE), Université libre de Bruxelles (ULB), Boulevard du Triomphe 2, 1050 Brussels, Belgium}
% ...

\emailAdd{collaboration@grand-observatory.org}

\abstract{The Giant Radio Array for Neutrino Detection (GRAND) is a proposed multi-messenger observatory of Ultra-High-Energy  (UHE) particles of cosmic origin.  Its main goal is to find the long-sought origin of UHE cosmic rays by detecting large numbers of them and the secondary particles created by their interactions like gamma rays and neutrinos. The GRAND Collaboration plans to achieve this using large arrays of radio antennas that look for the radio signals emitted by the air showers initiated by the interactions of the UHE particles in the atmosphere. Since 2023, three small-scale prototype GRAND arrays have been in operation:  GRAND@Nançay in France, GRAND@Auger in Argentina, and GRANDProto300 in China.  Together, their goal is to validate the detection principle of GRAND under prolonged field conditions, achieving efficient, autonomous radio-detection of air showers.  We describe the hardware, software, layout, and operation of the GRAND prototypes. Using their data, we show a first characterization of the local electromagnetic environment of each site and a measurement of the Galactic synchrotron emission. Despite challenges, the successful operation of the prototypes confirms that the GRAND instrumentation is apt to address the goals of the experiment and lays the groundwork for its ensuing stages.}


\maketitle
\flushbottom


\section{Introduction}

The Giant Radio Array for Neutrino Detection (GRAND)~\cite{GRAND:2018iaj} is a planned multi-messenger observatory of Ultra-High Energy (UHE) particles of cosmic origin---cosmic rays, neutrinos, and gamma rays. Its ultimate goal is to provide crucial evidence that addresses the long-standing question of the origin of UHE Cosmic Rays (UHECRs)~\cite{Anchordoqui:2018qom, AlvesBatista:2019tlv, Globus:2025tud}.  GRAND plans to do so, directly, by detecting a vast number of UHECRs and, indirectly, by detecting secondary gamma rays and UHE neutrinos produced from the interaction of UHECRs~\cite{Berezinsky:1969erk}. 
To achieve this, GRAND plans to use large arrays of radio antennas that search for distinctive radio frequency signals emitted by extensive air showers initiated by UHE particles interacting in the atmosphere~\cite{Huege:2016veh, Schroder:2016hrv}. 

The radio-detection of UHECRs is a mature technique. It has been demonstrated in past and present detectors using overground antenna arrays~\cite{Huege:2016veh, Schroder:2016hrv} by the Auger Engineering Radio Array (AERA)~\cite{PierreAuger:2012ker, PierreAuger:2025AERA}, the Low-Frequency Array (LOFAR)~\cite{LOFAR:2013jil, LOFAR:2014gtq}, the Taiwan Astroparticle Radiowave Observatory for Geo-synchrotron Emissions (TAROGE)~\cite{TAROGE:2022soh}, the Tianshan Radio Experiment for Neutrino Detection (TREND)~\cite{Charrier:2018fle}, and the Owens Valley Radio Observatory Long Wavelength Array (OVRO-LWA)~\cite{Monroe:2019zkp}; using in-ice arrays~\cite{Barwick:2022vqt} by the Askaryan Radio Array (ARA)~\cite{ARA:2015wxq} and the Antarctic Ross Ice Shelf Antenna Neutrino Array (ARIANNA)~\cite{ARIANNA:2014fsk}; and using balloon flights by the Antarctic Impulsive Transient Antenna (ANITA)~\cite{ANITA:2008mzi, ANITA:2010ect}. Concurrent with GRAND, the new Radio Detector of the Auger Prime upgrade~\cite{Horandel:2025xmh}, the upcoming Payload for Ultrahigh Energy Observations (PUEO)~\cite{PUEO:2020bnn}, and the Radio Neutrino Observatory in Greenland (RNO-G)~\cite{RNO-G:2024esr}, as well as the proposed next-generation Beamforming Elevated Array for COsmic Neutrinos (BEACON)~\cite{Southall:2022yil},  Radar Echo Telescope for Cosmic Rays (RET-CR)~\cite{RadarEchoTelescope:2021rca}, and POEMMA-Balloon with Radio (PBR)~\cite{PBR:2025fcw} also target the radio-detection of UHE particles, though adopting different, complementary~\cite{Wissel:2024rhp} detection strategies. 
References~\cite{MammenAbraham:2022xoc, Ackermann:2022rqc, Guepin:2022qpl} present an overview of the upcoming UHE detectors.

\begin{figure*}[t!]
 \centering
 \includegraphics[width=\textwidth, trim={0.6cm 0cm 2cm 0cm}, clip]{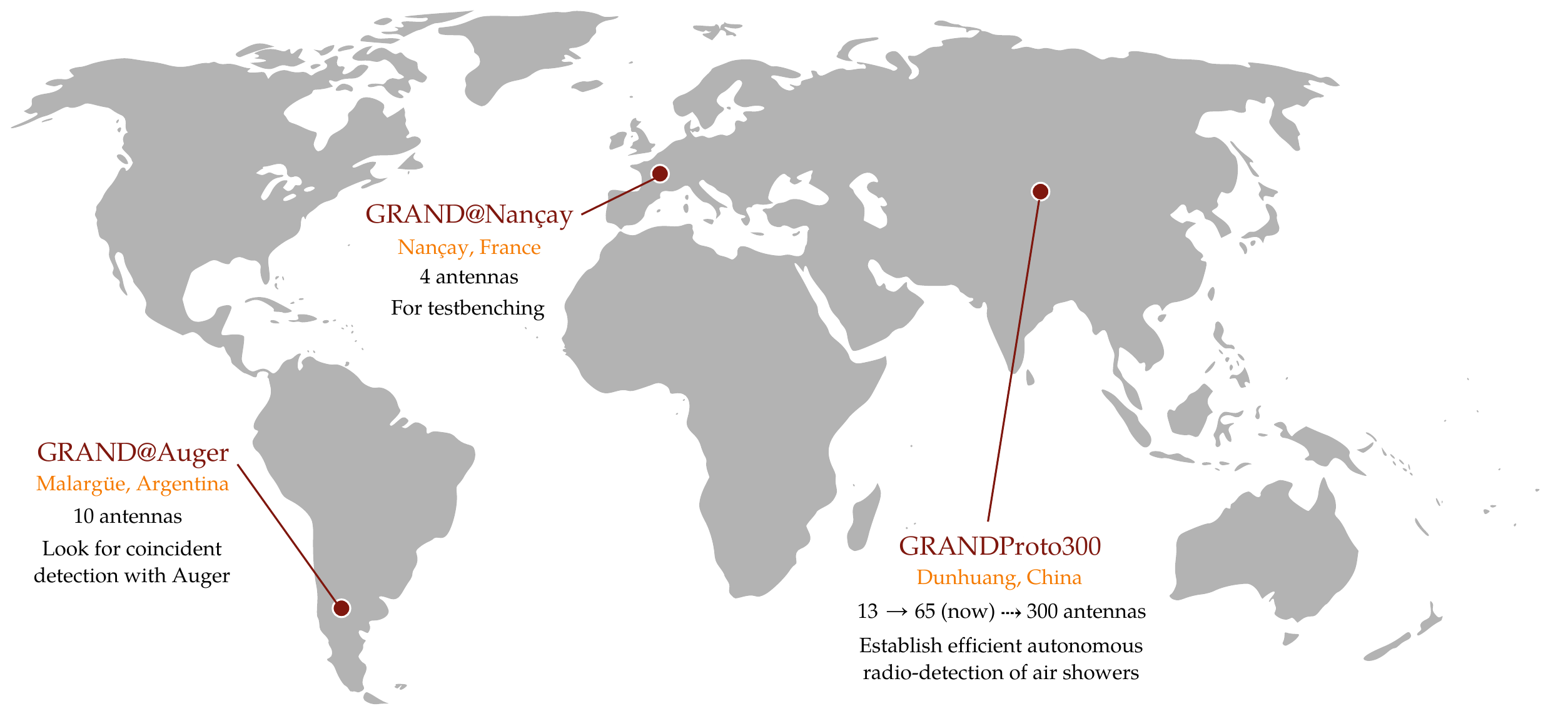}
 \caption{\textbf{The GRAND prototype arrays.} The three GRAND prototype arrays were deployed and have been in operation since 2023. They share a common basic instrumentation design, modified to suit the local environmental condition at the array site, and to test the performance of alternative hardware and software choices. This BlankMap-World-noborders.png has been obtained by the author(s) from the Wikimedia website,
where it is stated to have been released into the public domain. It is included within this article on that basis
(https://commons.wikimedia.org/wiki/File:BlankMap-World-noborders.png).}
 \label{fig:worldmap}
\end{figure*}

Building on the experience gained by these experiments---most directly, by AERA and TREND---GRAND plans to deploy antenna arrays at a larger scale~\cite{GRAND:2018iaj, Kotera:2024iyk}.  This involves increasing the number of deployed antennas---to eventually tens of thousands---and deploying them in arrays that are sparser than before---with km-scale antenna separations.  This up-scaling is required for the antenna arrays to monitor the large volumes of the atmosphere needed, creating an aperture of 107000\,km$^2$sr, which delives vast numbers of detected UHECRs---6500 UHECRs above $10^{19.5}$\,eV per year~\cite{GRAND:2018iaj}---and a sensitivity to even tiny fluxes of UHE neutrinos---2 orders of magnitude below the IceCube sensitivity~\cite{GRAND:2018iaj, Moller:2018isk, Ackermann:2022rqc, MammenAbraham:2022xoc, Guepin:2022qpl}.

We implemented a staged construction strategy for GRAND~\cite{GRAND:2018iaj, deMelloNeto:2023zvk, Martineau-Huynh:2025hkr} intended to validate the design and operation of the experiment on smaller scales before deploying larger arrays. Figure~\ref{fig:worldmap} shows the locations of the GRAND prototypes, in operation since 2023~\cite{Martineau-Huynh:2025hkr}: \textbf{GRAND@Nançay} in Nançay, France, \textbf{GRANDProto300}~\cite{Ma:2025xgd} in Dunhuang, China, and \textbf{GRAND\allowbreak@Auger}~\cite{deErrico:2025qsa} in Malargüe, Argentina, at the site of the Pierre Auger Observatory (Pierre Auger)~\cite{PierreAuger:2015eyc}.  

Their purpose is to validate two main requirements that GRAND must fulfill to achieve its goals. First, the prototypes must validate the viability of autonomous radio-detection of air showers in a sparse radio array~\cite{Correa:2023maq, LeCoz:2023bie, Correa:2024rxi,  Correa:2025ovu}, unaided by particle detectors, and despite the presence of stochastic noise and anthropogenic background that may mimic the electromagnetic signatures expected from showers. Second, the prototypes must validate that the arrays are able to operate under harsh environmental field conditions---high temperature, sandstorms, rainstorms---for prolonged periods of time. The lessons gained from the prototypes inform the design of the future stages of the experiment.

In this paper, we introduce the design and operation of the GRAND prototype arrays, focusing on the two larger ones, GRANDProto300 (GP300) and GRAND@Auger (G@A), and we present a first look at their data. The remainder of the paper is organized as follows.  Section~\ref{sec:layout} introduces the GRAND prototype arrays, their sites, and layouts.  Section~\ref{sec:setUp} describes the GRAND detection unit, its hardware, and performance.  Section~\ref{sec:trigger_DAQ} presents the self-trigger algorithm and the software processes that comprise the GRAND data acquisition chain. Section~\ref{sec:data_management} describes how the collected data are managed, \textit{i.e.}, the data transfer pipeline from local and central storage on-site to permanent storage off-site, and the data-quality monitoring tools of the experiment. Section~\ref{sec:spectra} shows the first characterization of the local electromagnetic environment and the measurement of the Galactic synchrotron emission using data from the GRAND prototypes. Section~\ref{sec:ending} summarizes and presents the perspectives.


\section{Overview of the GRAND prototype arrays}
\label{sec:layout}

\begin{figure*}[t!]
 \centering
 \includegraphics[width=\textwidth]{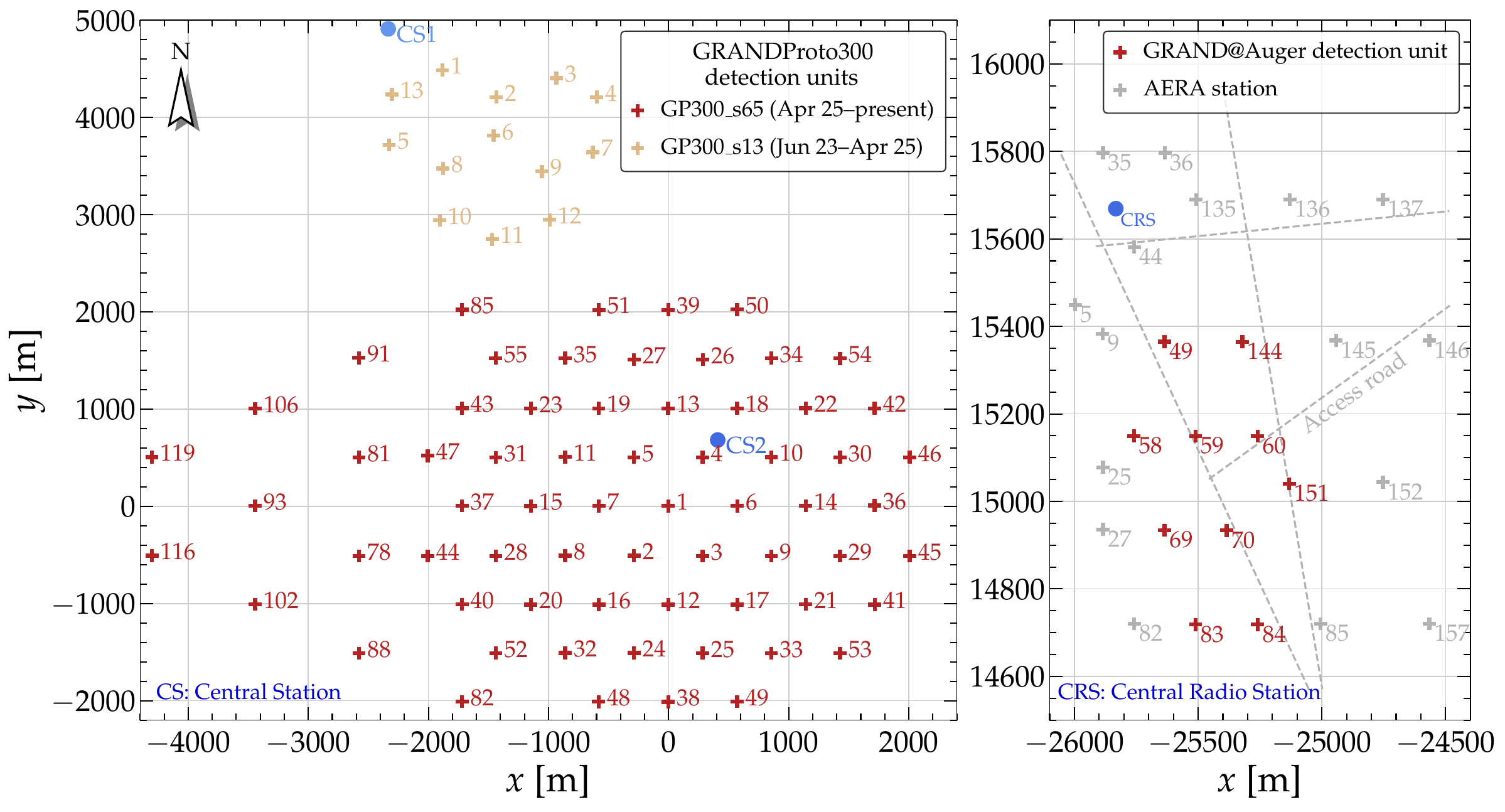}
 \caption{\textbf{Array layout of GRANDProto300 ({\it left}) and of GRAND@Auger ({\it right}).} Each GRAND Detection Unit (DU) has an assigned identification number.  The Central Station in GRANDProto300 and the Central Radio Station in GRAND@Auger house the central data acquisition (DAQ) systems of the experiments. Positions for GRANDProto300 are given with respect to DU 1 of GP300\_s65.  For GRAND@Auger, DU positions are given in the Auger Coordinate System. We also show the nearby AERA radio-stations (in grey) and access roads.}
 \label{fig:arrays}
\end{figure*}

The site and array layout for each GRAND prototype were decided to fulfill their specific goals: testbenching for GRAND@Nançay, demonstrating the autonomous radio-detection of air showers for GP300, and observing coincident detection of air showers with the Pierre Auger  detectors in G@A.  The radio-frequency electromagnetic environment varies between sites, with the anthropogenic background at GRAND@Nançay being the loudest. For GP300, the background is appreciably smaller than at G@A, predominantly due to GP300 being farther from human activity.

\smallskip

\textbf{\textit{GRAND@Nançay}} is a small array of four GRAND Detection Units (DUs) located at the Nançay Radio Observatory, in France. It was conceived as a field site accessible to European laboratories, facilitating testbenching of equipment and new ideas in design, trigger, and data acquisition.  The design of GRAND@Nançay resembles that of the other two prototype arrays, though with differences, such as having antennas of lower height and communications via optical fiber instead of wireless. Its small size is sufficient for its purpose as a testbed.  In the remainder of the paper, we focus on GRANDProto300 and GRAND@Auger.

\smallskip

\textbf{\textit{GRANDProto300}} is the GRAND pathfinder array in the Northern Hemisphere.  In its final stage, it will be a 300-antenna array; currently, it exists in partial configurations.

The 300-antenna stage of GP300 will span an area of about 200~km$^2$ in the Gobi Desert near Xiaodushan, in northern Dunhuang, China, making it the largest of the three prototypes. The site is a largely radio-quiet region in the 50--200~MHz radio-frequency band, which GRAND adopts as its working frequency band. The precise location was chosen after a survey of regions that contain a relatively flat area capable of hosting the final 300-antenna stage of GP300, and that was located near a mountain range that acts as an additional neutrino interaction target, as per the GRAND science case~\cite{GRAND:2018iaj}.

In June 2023, the deployment of the 13-antenna stage, GP300\_s13, was completed~\cite{Ma:2023siw}. During its nearly two years of operation, it enabled the validation of the GRAND detection concept and the improvement of the system design. By April 2025, the array was extended to its current 65-antenna stage, GP300\_s65, which has been operating since, and whose additional goals include testing the efficiency and purity of the autonomous radio-detection of inclined air showers from UHECRs.  GP300, even in partial configuration, is expected to detect UHECRs with energies between 100\,PeV and a few EeV, probing the Galactic to extragalactic transition region of cosmic ray origin.

Figure~\ref{fig:arrays} shows the layout of GP300\_s13 and GP300\_s65. In both GP300 and G@A, DUs are arranged in a hexagonal layout, a choice made to optimize the simulated efficiency of air-shower detection with respect to the density of antennas on the ground.  A hexagonal array has fewer directions along which the performance of air shower-reconstruction methods is degraded as a result of the shower trajectory being aligned with the axes of symmetry of the array layout.

GP300\_s13 had a three-hexagon layout, covering an area of about 2~km$^2$ with a DU spacing of about 500~m. The Central Station 1 (CS1) was built north-west of the array; it consisted of three sets of container rooms that stored the central data acquisition (DAQ) computer and the lodging and cooking facilities. GP300\_s13 was decommissioned in April 2025 and its DUs were incorporated into GP300\_s65, which has remained operational since. GP300\_s65 was built south of GP13, covering an area of about 30~km$^2$. The array has a 55-DU dense region to the east, with a DU spacing of 577~m, and a 10-DU sparse region to the west, with a spacing of 1000~m. A new Central Station (CS2) was built for GP300\_s65 and placed in its center.  

Preliminary simulations~\cite{Kato:2025cmc} estimate that the full GP300 array may detect about 130 inclined cosmic-ray events per day---for an ideal 100\% uptime of the detector---in the energy range of $10^{17}$--$10^{20}$~eV, reaching a trigger efficiency of 90\% already at $10^{18.3}$~eV. These estimates account for the processing of the radio signal from the air shower through the Radio-Frequency (RF) chain of the detector units and for the persistent noise from the Galactic emission. Improvements are underway to optimize the cosmic-ray trigger rate, suppress the electromagnetic noise and background, and refine the selection of cosmic-ray candidates~\cite{Lavoisier:2025ase, Guelfand:2025hqr, Ferriere:2025csu, Gulzow:2025hyc, Benoit-Levy:2025rcq, Zhang:2025ssu}.  Because GP300\_s13 and GP300\_s65 are smaller, they are sensitive to the lower end of the cosmic-ray energy range, between about $10^{16.5}$~eV and $10^{18.5}$~eV, where the cosmic-ray flux is higher.  The search for these cosmic rays has already yielded its first likely candidates in GP300\_s65 data~\cite{Lavoisier:2025ase, Guelfand:2025hqr, Ferriere:2025csu}.

\smallskip

\textbf{\textit{GRAND@Auger}} results from an agreement between the GRAND and Pierre Auger Collaborations to repurpose 10 AERA Phase-2 radio-stations in Malargüe, Argentina, into GRAND DUs. The G@A array is embedded within the region of the Pierre Auger detector array, consisting of water-Cherenkov particle detectors (Surface Detectors---SD), fluorescence telescopes, and radio antennas~\cite{Horandel:2025xmh}. This offers, uniquely among the three GRAND prototypes, the opportunity to validate the radio-detection---and the ensuing reconstruction---of air showers that are simultaneously detected by both the Pierre Auger Observatory and the GRAND@Auger antennas. 

The deployment of G@A was completed in August 2023 and it has been operational since then. Figure~\ref{fig:arrays}, right, shows the layout of G@A. The GRAND prototype was deployed within the AERA array, itself located within the Pierre Auger 750-m SD array, and within the field of view of the Coihueco and HEAT fluorescence detectors~\cite{PierreAuger:2009esk, Mathes:2011zz}. The G@A DUs are laid on two superimposed hexagons with a predominant spacing of 250~m, covering an area of 0.5~km$^2$. The central DAQ computer is located in the AERA Central Radio Station (CRS). Nearby roads facilitate access to the site.

For G@A, using a denser detector layout enhances the detection of less energetic and less inclined events.  The estimated event rate for the Pierre Auger 750-m SD array is $390 \pm 17$ showers per day over 25~km$^2$~\cite{Maris:2011zz}. This corresponds to about 8 showers per day detected by the Pierre Auger SD array within the area of G@A, making it viable to search for showers measured simultaneously by G@A and Pierre Auger detector arrays. One such coincident event has been detected so far~\cite{deErrico:2025qsa}. Because the geomagnetic field is weaker at the G@A site than at the GP300 site, a shower with the same number of particles will emit a weaker radio-frequency signal ({\it i.e.}, a smaller-amplitude electric field) in G@A than at GP300. This leads to weaker signals in the antennas, making it generally more challenging to detect showers in G@A.

\begin{figure}[t!]
 \center
 \includegraphics[height=7cm]{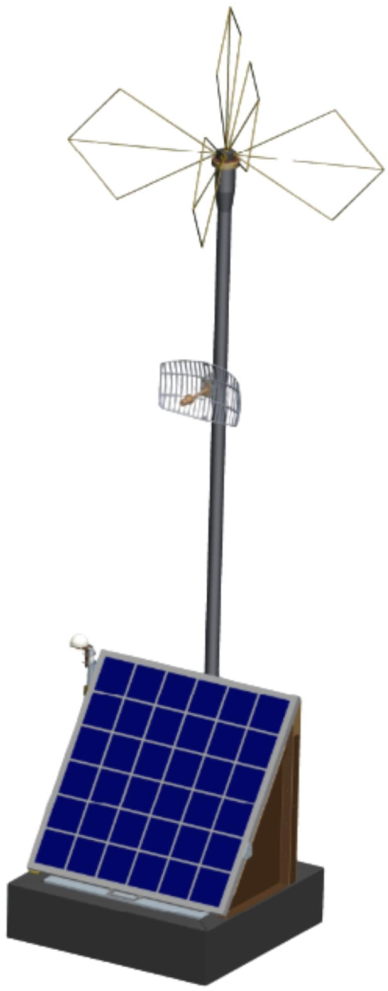}
 \hspace*{0.6cm} 
 \includegraphics[height=7cm]{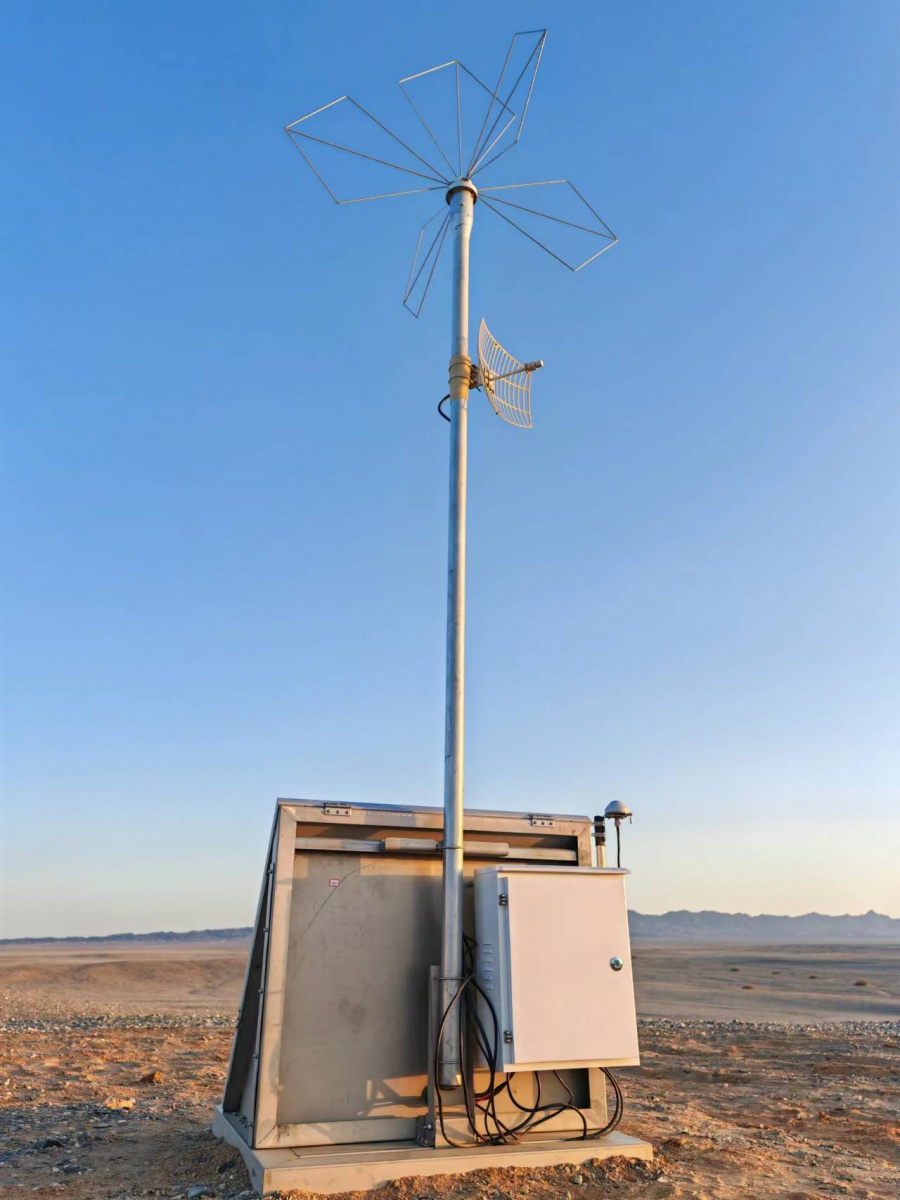} 
 \caption{\textbf{GRAND detection unit deployed in GRANDProto300.} \textit{Left:} The 3.5-m vertical pole holds the antenna and the nut (Section~\ref{sec:setUp_mechanical}) containing the low-noise amplifier (Section~\ref{sec:setUp_antenna_lna}). Midway along the pole lies the communications antenna.  The solar panel at the bottom also functions as a door, providing access to an enclosure that houses the battery and charge controller. \textit{Right:} Photograph of a unit deployed at the GP300 site, in China. The box attached at the bottom of the pole stores the front-end board (Section~\ref{sec:FEB}) and on its top-right corner is the GPS antenna.}
 \label{fig:gp13_du}
\end{figure}

\begin{figure}[t!]
\center
\includegraphics[height=7cm]{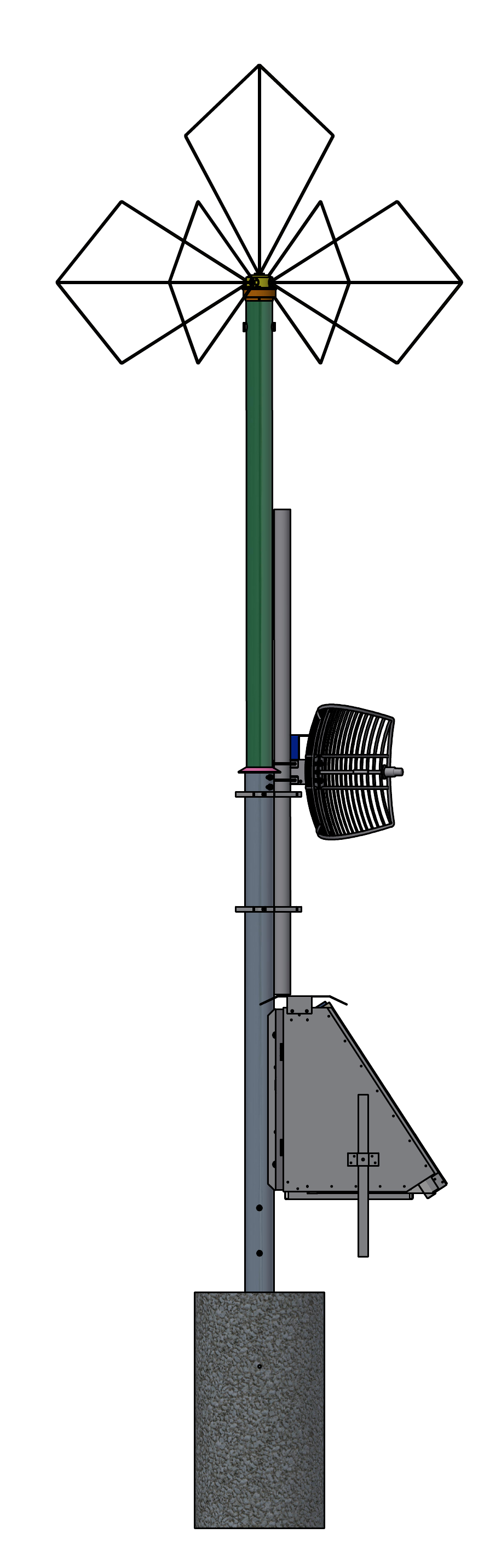}
\hspace*{0.6cm} 
\includegraphics[height=7cm]{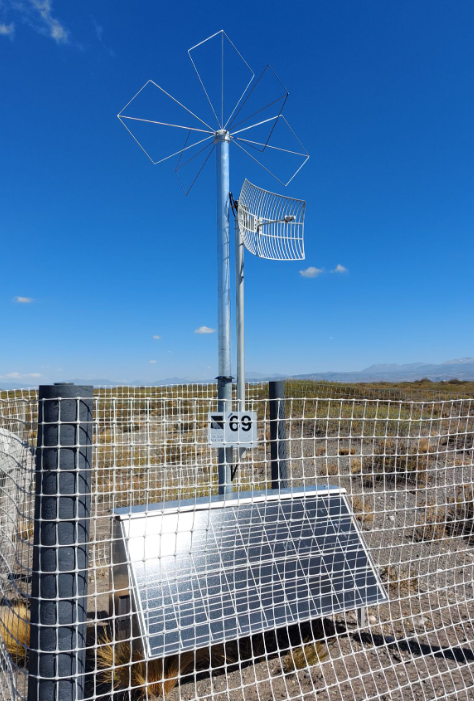} 
\caption{\textbf{GRAND detection unit deployed in GRAND@Auger.} 
 \textit{Left:} The green pole is an aluminum tube attached to the original AERA pole with an adapter ring to increase the antenna height to 3~m (Section~\ref{sec:setUp_mechanical}). The antenna head, containing the low-noise amplifier (Section~\ref{sec:setUp_antenna_lna}), is connected to a sleeve on the top of the pole, which is kept in place with a bolt. A communications antenna is mounted midway along the pole. The triangular shaped box in the lower half, from the original AERA design, contains the battery, the front-end board (Section~\ref{sec:FEB}), and the charge controller. It also serves as support for the solar panel. \textit{Right:} Photograph of a unit deployed at the G@A site, in Argentina. The plastic mesh around the DU protects it from free-roaming animals in the area.}
\label{fig:gata_du}
\end{figure}


\section{The GRAND detection unit}
\label{sec:setUp}

The GRAND detection unit (DU) consists of an antenna---the Horizon Antenna---and the electronics to process, store, and communicate the measured signals. The DU design is optimized to detect radio-frequency signals that reach the array from near-horizontal directions, as expected from inclined air showers initiated by cosmic rays and neutrinos~\cite{Huege:2016veh, Schroder:2016hrv}.  

The GRAND DU design has built-in flexibility, allowing it to be modified and adapted if needed. For instance, this feature was exploited in G@A when the original DU design was altered to accommodate it to the existent mechanical structure of AERA. In addition, the different environmental conditions at each site required specific instrumentation adjustments. Site-specific adaptations include heat and moisture countermeasures, hibernation conditions, and tailored trigger settings. We also profited on this opportunity to assess the performance of alternative hardware and software choices and thus field-test options for later stages of GRAND.


\subsection{Mechanical structure}
\label{sec:setUp_mechanical}

Figure~\ref{fig:gp13_du} shows the mechanical structure of the DUs deployed at GP300. A 3.5-m pole holds a ``nut'' onto which the five arms of an antenna are mounted. The nut houses the Low-Noise Amplifiers (LNAs). In total, three cables run from the LNAs through the main pole to the GRAND Front-End Board (FEB), which is kept inside a box within a ventilation casing on the back of the triangular-shaped base on the ground. A mesh communications antenna serving WiFi communication with the corresponding central station is mounted about halfway on the main pole. The power for all elements of the DU is provided by a solar panel connected through a charge controller to a battery. The solar panel functions as a door that allows access to both. In addition, the pole can be lowered to facilitate access to the nut. 

Figure \ref{fig:gata_du} shows the mechanical structure of the DUs deployed in G@A, closely resembling those in GP300, modified to better fit the original AERA stations. In particular, the smaller vertical pole was extended by fitting a hollow aluminum tube inside the existing pole, thus raising the antenna to a height of 3~m---50~cm lower than the antennas at GP300. Raising the antenna 3.5~m above ground may have caused vibrations that the setup could not sustain. The original AERA solar panel and the metal enclosure were kept, with the GRAND FEB box, battery, and charge controller stored inside. In the following sections, we discuss the characteristics of the antenna, the LNAs, the GRAND FEB, power supply, and communications setup.


\subsection{Antenna and Low-Noise Amplifier}
\label{sec:setUp_antenna_lna}

\begin{figure}[t!]
 \centering
 \vspace*{-0.5cm}
 \raisebox{1.2cm}{\includegraphics[width=.49\columnwidth, trim={0 0 0 0.8cm}, clip]{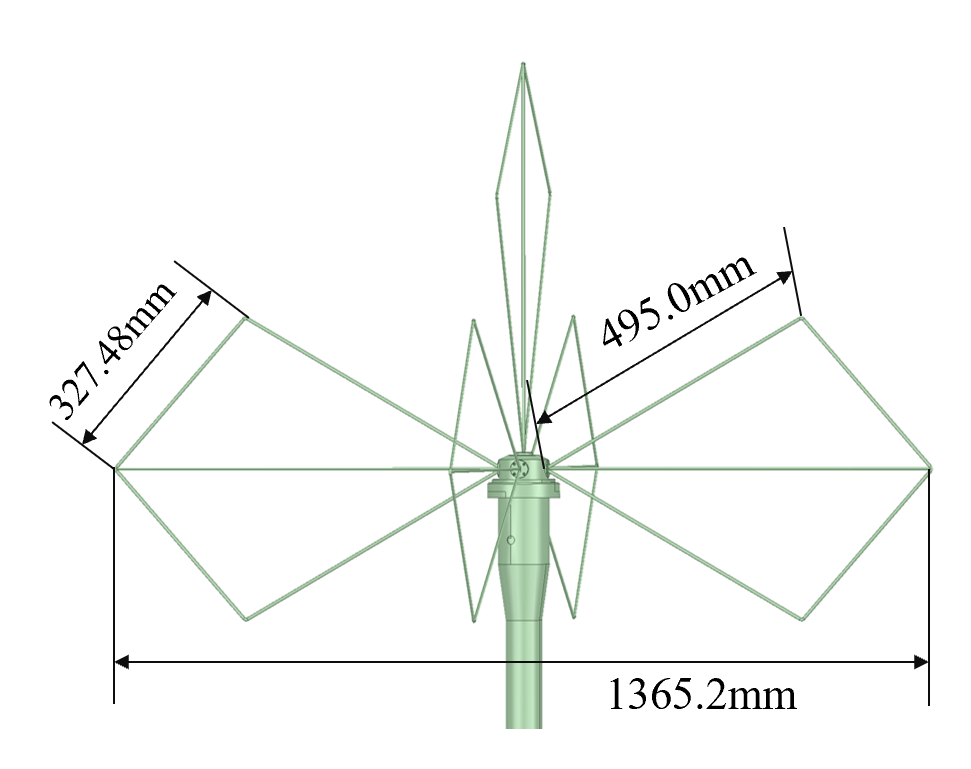}}
 \includegraphics[width=0.3\columnwidth]{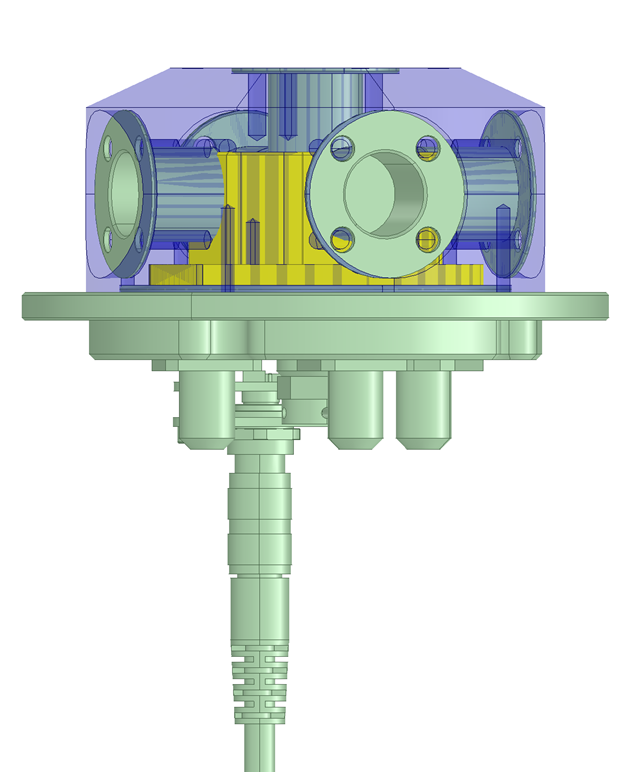}
 \caption{\textbf{Schematic diagrams of the GRAND Horizon Antenna and nut.}  \textit{Left:} The antenna, with three arms: two dipoles in the North-South and East-West directions, and a monopole in the vertical one. \textit{Right:} The nut: the acrylic, in blue, holds the antenna arms and protects the interior metal connectors, inside the support, in yellow. These connect the antenna arms to the LNA board, held at the metal base, in green.}
 \label{fig:antenna_arms_nut}
\end{figure}

\textbf{\textit{Antenna.---}}Figure \ref{fig:antenna_arms_nut}, left, shows that the GRAND Horizon Antenna consists of five arms: two dipoles in the North-South and East-West directions, and a monopole in the vertical direction. The dipole size is roughly 1.4~m, to match the GRAND working frequency band of 50--200~MHz. Figure~\ref{fig:antenna_arms_nut}, right, shows the acrylic nut with an internal connector to which the antenna arms are attached. The LNA board is housed in the metal base of the nut and is powered by a coaxial cable that also transports received signals to the FEB. This structure is common to both prototype arrays.

\begin{figure}
    \centering
    \includegraphics[width=\linewidth, trim={8.5cm 28.cm 11cm 6.3cm}, clip]{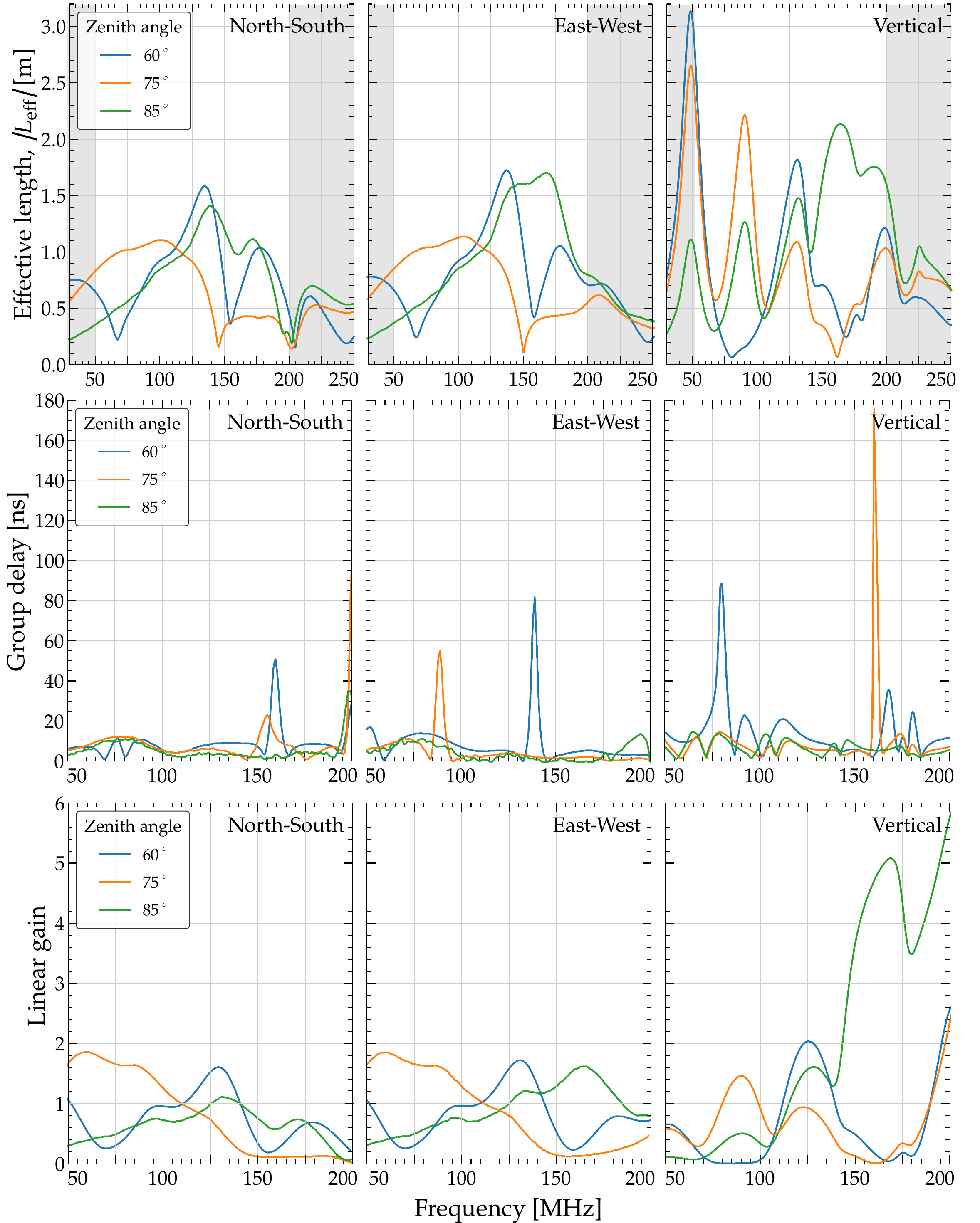}
    \caption{\textbf{Simulated Vector Effective Length (VEL)} (\textit{top}), \textbf{group delay} (\textit{middle}) \textbf{and linear gain }(\textit{bottom}) \textbf{of the GRAND Horizon Antenna.} Each property is shown as a function of frequency for the three antenna polarizations: North-South (\textit{left}), East-West (\textit{center}), and vertical (\textit{right}). We show results for an azimuth angle of $45^\circ$ measured counter-clockwise from the North, and for three illustrative choices of the zenith angle: $60^\circ$, $75^\circ$, and $85^\circ$.}
    \label{fig:ant_prop}
\end{figure}


Figure \ref{fig:ant_prop}, top, shows the simulated response of the GRAND Horizon Antenna to an incoming radio-frequency signal, expressed in the form of the magnitude of the Vector Effective Length (VEL), $\mathbf{L}_{\rm eff}$. The VEL is the transfer function between the incoming electric field, $\mathbf{E}$, and the open-circuit voltage at the antenna terminals, $V_{\rm oc}$, {\it i.e.}, 
\begin{equation}
    V_{\rm oc} = \mathbf{L}_{\rm eff} \cdot \mathbf{E}.
\end{equation}
\noindent The larger the magnitude of the VEL, the higher the voltage generated by the antenna for a given incident electric field, provided that the electric field polarization and the VEL are aligned.

For the Horizon Antenna, the VEL was simulated using the Ansys HFSS (High-Frequency Structure Simulator) program, with a setup that included the complete mechanical structure of the DU. The resulting VEL is an elaborate function of frequency, zenith, and azimuth angle that fully characterizes the antenna response to an arbitrary incoming plane-wave signal. This is essential for producing simulated measurements and interpreting real ones. For the Horizon Antenna, Fig.~\ref{fig:ant_prop}, top, shows that the North-South and East-West arms have comparable VELs, expected given the dipoles are of equal size. The vertical arm differs likely due to the different size of the monopole, its different orientation, and the reflection of radio-frequency waves off the ground.

Given the complexity of the VEL of the Horizon Antenna, its validation in the field remains to be completed. A calibration campaign is planned to verify our simulations. It will be performed by sending radio-frequency waves of frequencies between 50-200 MHz, with a spacing of 1 MHz, to the GRAND Horizon Antennas. The emitting antenna will have a known location and will be movable (by truck or drone). In this way, we will have the antenna response dependency on frequency, azimuth, and zenith angles fully characterized. A comparison with the Galactic noise measurement will confirm the absolute value of the calibration up to ~10\%. This will be an improvement over a simulation-only estimate of about a factor of three. 

Figure~\ref{fig:ant_prop}, middle, shows the group delay of the Horizon Antenna, {\it i.e.}, the derivative with respect to the frequency of the phase shift of the DU transfer function. A varying group delay throughout the working frequency range can lead to signal dispersion and loss of amplitude due to interference effects. For air-shower radio-detection and reconstruction, the shape and amplitude of the signal are important assets used, for example, in triggering algorithms. For the Horizon Antenna, the group delay is nearly flat in frequency for the three antenna polarizations, with differences well below 20~ns that do not significantly distort the expected pulse shape. Given the frequency bandwidth of 200~MHz, group delay variations smaller than its inverse (5~ns) are acceptable. Although there are sharp peaks in the group delay, they are narrow enough to have only a small effect on signal power. The antenna response is embedded in the {\sc GRANDlib} processing pipeline~\cite{GRAND:2024atu} and is used in the reconstruction of the properties of the air showers detected by GRAND.

Figure~\ref{fig:ant_prop}, bottom, shows the gain of the Horizon Antenna. The gain of an antenna measures the sensitivity of the antenna in a specific direction with respect to its average sensitivity. It is the ratio between the power transmitted to the antenna feedpoint by a signal coming from one specific direction and the same quantity averaged over all directions. Figure~\ref{fig:ant_prop}, bottom, shows it for an incident azimuth angle of $45^{\circ}$ to facilitate comparison between the horizontal North-South and East-West polarizations. These exhibit similar behavior, with an enhanced sensitivity below 100\,MHz, where most of the radio-frequency signal from extensive air showers is expected, for zenith angles of 60--$75^{\circ}$. At a zenith angle of $85^{\circ}$, the gain remains close to unity for a large part of the 50--200\,MHz bandwidth, despite stronger ground attenuation. The slight difference between the North-South and East-West polarizations is due to the influence of the solar panel placed at the foot of the antenna. Similarly as for the VEL and the group delay, also depicted in Fig. \ref{fig:ant_prop}, the Vertical polarization exhibits a different behavior. The gain increases with the zenith angle, as expected, but also with frequency, a feature we understand as a consequence of ground reflections.

\begin{figure}[t!]
 \centering
 \includegraphics[width=.8\textwidth]{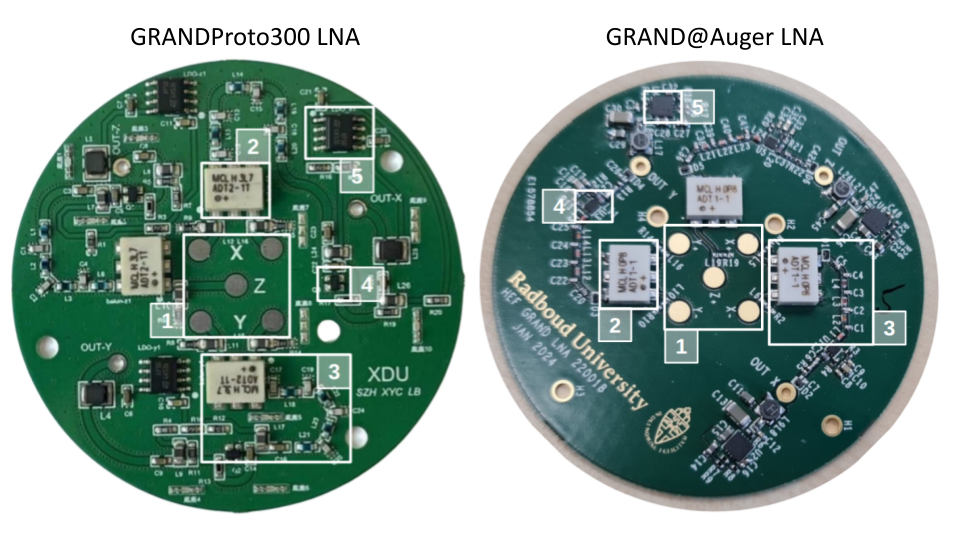} 
 \caption{\textbf{Low-Noise Amplifier (LNA) board deployed at GRANDProto300} (\textit{left}) \textbf{and GRAND@Auger} (\textit{right}). The main components of the boards are: (1) connectors to the antenna, (2) balun, (3) matching network, (4) amplifier, and (5) power.  Although the components are in similar locations, their specifications are not the same in both boards.}
 \label{fig:lna_pcb}
\end{figure}

\smallskip

\textbf{\textit{LNA.---}}Figure~\ref{fig:lna_pcb} shows the two designs of the LNA boards deployed in GP300 and G@A. For GP300, the LNA board was designed to provide a stable gain of about 20~dB in the range of 50--200~MHz. As shown in Fig.~\ref{fig:lna_pcb}, left, in the GP300 board the X and Y inputs are connected to the dipoles and the Z input, to the monopole. When powered with 5~V from the FEB, the LNA board draws 47~mA for each channel, which leads to a power consumption of about 235~mW for the whole board. An electrostatic discharge protection structure guards the LNA from damage due to static electricity. The balun converts differential signals into single-ended signals and performs impedance transformation. The matching network, located at the front-end of the LNA, converts the impedance to 50~$\Omega$, minimizing impedance matching issues at the LNA input port. 

For G@A, the LNA design was focused on providing higher gain at a low power consumption. The design delivers a stable gain of about 30~dB for each of the X, Y, Z inputs, in the same range as GP300, 50--200~MHz. Like in GP300, the board is powered with 5~V by the GRAND FEB; each channel draws 40~mA and the total power consumption of the board is  200~mW. As shown in Fig.~\ref{fig:lna_pcb}, right, the path lengths for the X, Y, and Z inputs in the LNA board are exactly the same, leading to equal signal time delays for all channels. This allows for the study of the time dependence of the polarization in an unbiased way, avoiding additional software corrections to the timing of the individual polarization directions. Given the weather conditions in Argentina, we implemented moisture countermeasures to prevent water damage: the nuts are sealed with an additional layer of silicone, and the LNA boards are coated with transparent lacquer.

The main differences in the output signals provided by the two designs of the LNA board are the gain and the matching network. For GP300, the LNA has uniform matching across the full frequency range. For the G@A model, due to the high rate of anthropogenic Radio Frequency Interferences (RFIs) in the low frequency range, its matching is optimized for higher frequencies, {\it i.e.}, above 80 MHz. With this design choice, we avoid saturation of the LNA by reducing the gain at the low frequency end while keeping the relative phase delay small. The difference in maximum gain for the LNA boards is compensated for by the amplifier in the GRAND FEB, as described in the next section.


\subsection{Front-End Board}
\label{sec:FEB}

\begin{figure*}[t!]
\center
\includegraphics[width=\textwidth]{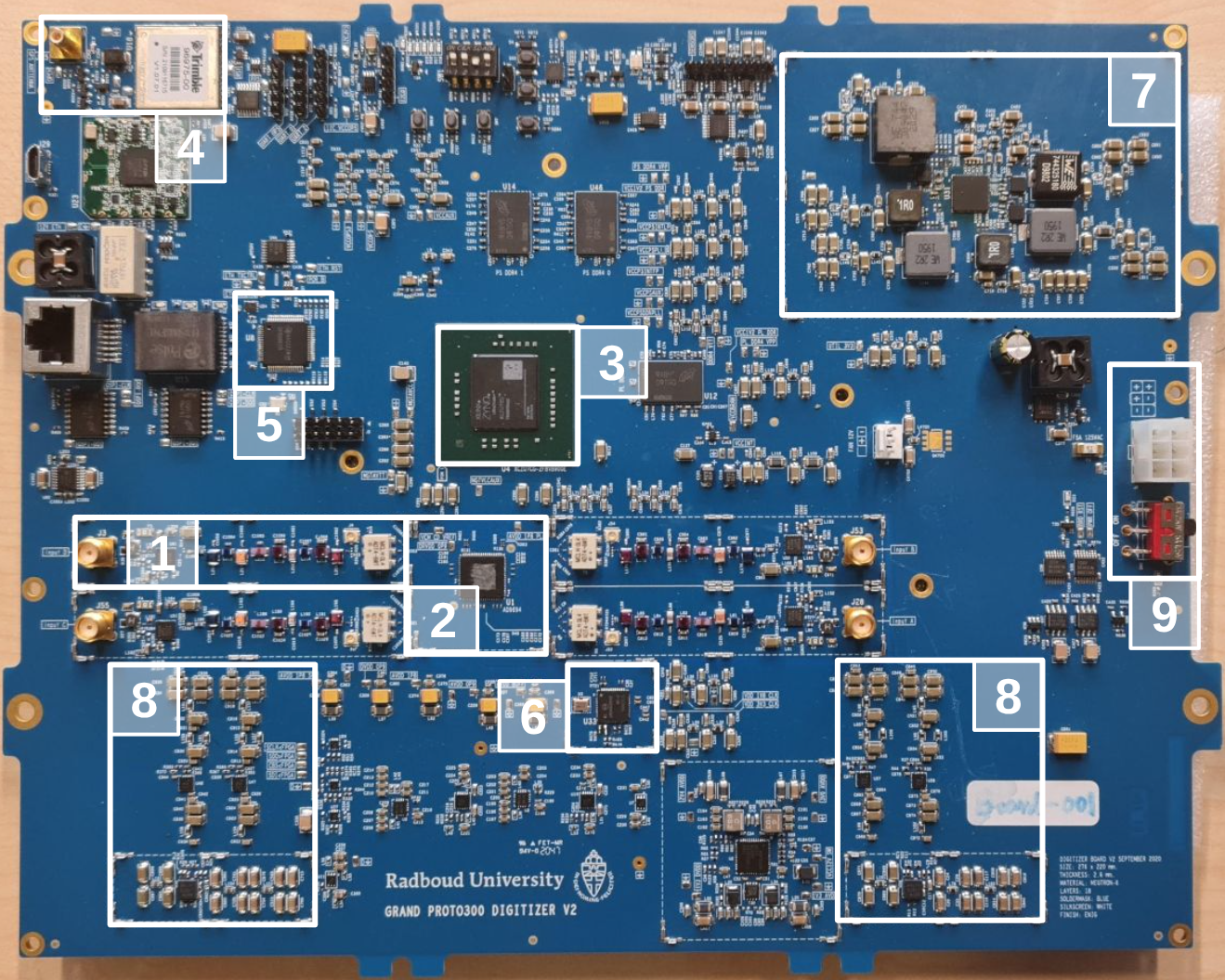} 
\caption{\textbf{The GRAND Front-End Board (FEB).} The main components: (1) signal input and filter chain, with four inputs available; (2) analog-to-digital (ADC) chip; (3) system on chip: field-programmable gate array (FPGA) and central processing unit (CPU) ; (4) Global Positioning System (GPS) chip and connector; (5) Ethernet chip; (6) clock; (7) power supply for the digital part of the board; (8) power supply for the analog part of the board; and (9) power connector and switch.}
\label{fig:Eboard}
\end{figure*}

\begin{figure}[t!]
 \center
 \includegraphics[width=0.5\columnwidth,trim={.5cm .6cm 0cm .5cm}, clip]{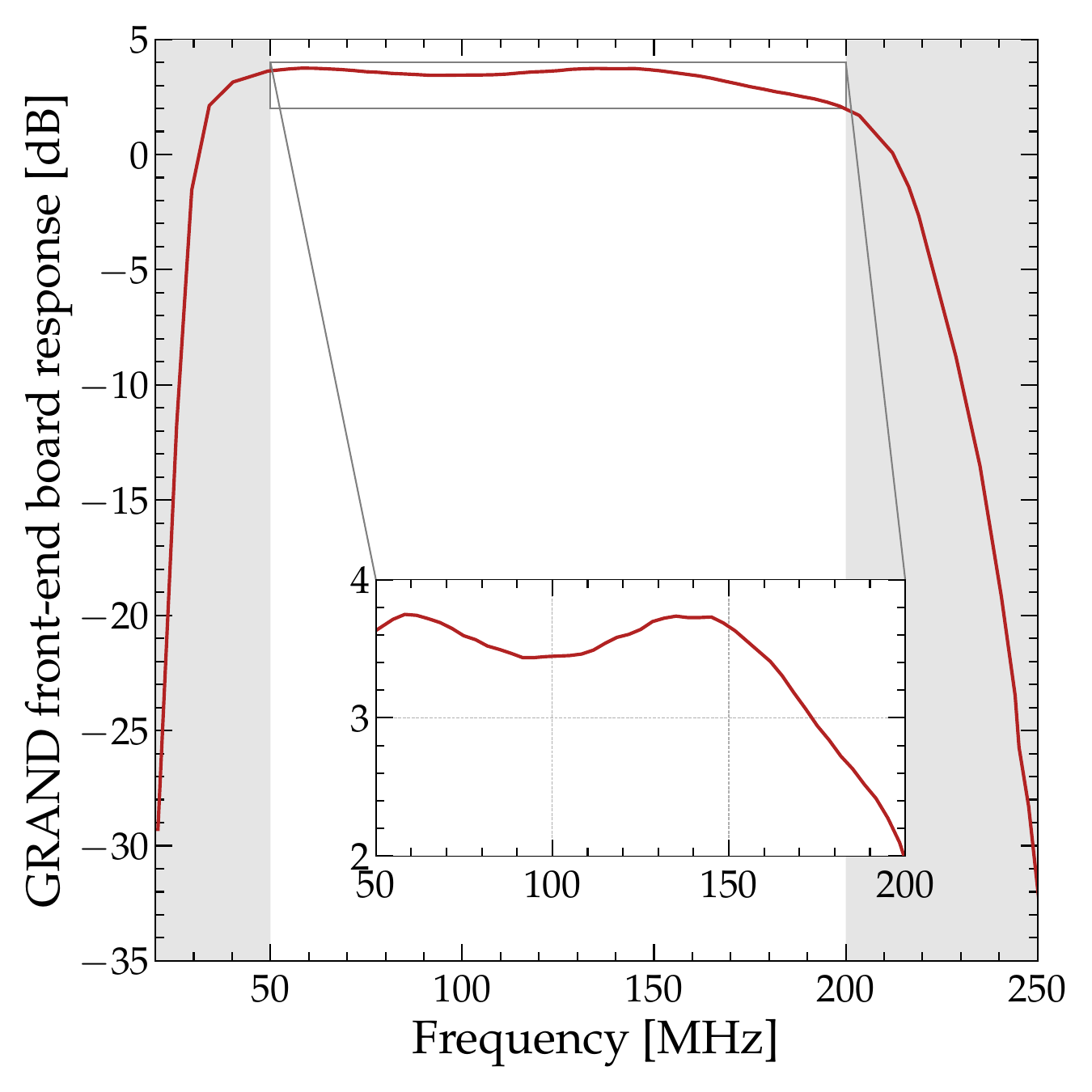} 
 \caption{\textbf{Measured response of the analog filtering on the prototype GRAND FEB with a set 4 dB VGA gain.} 
 The plateau at around 4~dB indicates that the signal is properly amplified within the GRAND working frequency range of 50--200 MHz.}   
 \label{fig:BPass}
\end{figure}

Figure~\ref{fig:Eboard} shows the GRAND FEB prototype used in GP300 and G@A. We describe its primary components below.

\smallskip

\textbf{\textit{Analog chain.---}}The analog chain, component (1) in Fig.~\ref{fig:Eboard}, combines several functionalities. The analog signals, coming from the LNA, are first shaped through a fifth-order elliptical 30--200\,MHz-bandpass filter. Then, they pass through the Variable Gain Amplifier (VGA), whose setting is controlled by the digital part of the board. The gain can vary between -10 and +20\,dB. Figure~\ref{fig:BPass} shows the measured analog response of the FEB board for a set VGA gain of 4~dB. In the GRAND working frequency range of 50--200~MHz, a plateau is observed at the expected 4~dB, with frequency dependent losses of at most half a dB. In GP300, we use a VGA gain of 20\,dB; in G@A, we use a 0\,dB gain. This choice compensates for the differences in the LNA gain and ensures that the final signal output has a similar total gain for both prototype arrays. The FEB has four analog chain signal inputs, one for each antenna polarization, and a spare that typically remains unused. The analog chain also supplies power to the front-end LNA boards through a bias-T.

\smallskip
 
\textbf{\textit{ADC chip.---}}The FEB digitizes the signals after they pass the analog chain. The digitization uses a 14-bit 500\,MSPS, 4-channel Analog-to-Digital Converter (ADC; Analog Devices, AD9694), component (2) in Fig.~\ref{fig:Eboard}, in differential mode with an input voltage range of $\pm 0.9$\,V. This setting allows us to measure the background noise at around 30 ADC counts, leaving a dynamic range of about 9 effective bits. This yields a compromise between accurate noise determination and maximum signal detection that allows us to use the variations in the Galactic noise as a monitoring and calibration tool. The 500 MHz clock used by the ADC is generated locally on the board. It is not synchronized between different DUs.

\smallskip

\textbf{\textit{FPGA and CPU chip.---}}The digitized signals are moved into a System on a Chip (SoC) (Xilinx, Zynq Ultrascale+, XCZU7CG-2FBVB900E), component (3) in Fig.~\ref{fig:Eboard}. It consists of a Field-Programmable Gate Array (FPGA) combined with four embedded hardcore Central Processing Units (CPUs), two ARM-A53 and two ARM-R5. The SoC performs tasks such as event triggering, event building, communication, and buffering. These functions are divided into tasks performed by the firmware that runs on the FPGA, and tasks performed by the Local Data Acquisition (DAQ) software running in the Petalinux (embedded Linux system) in the CPU---described in Section \ref{sec:trigger_DAQ}. Double Data Rate (DDR) memory, with a total size of 512\,MB, is available to be used by the FPGA and CPU chip.

\smallskip

\textbf{\textit{GPS and data transfer.---}} Timing and temperature information is obtained from the Trimble RES SMT360 GPS chip, component (4) in Fig.\ref{fig:Eboard}, which has an accuracy of 15\,ns. Wireless data transfer between the Detection Unit (DU) and the Data Acquisition (DAQ) computer is performed through the Ethernet chip, component (5) in Fig.~\ref{fig:Eboard}.

\smallskip

\textbf{\textit{The FEB box.---}}As described in Section \ref{sec:setUp_mechanical}, the FEB is stored in a sealed metal box. The FEB box also acts as a Faraday cage, preventing the electronic noise generated by the FEB components from interfering with the measurements. Due to the high temperatures during the summer at the GP300 and G@A sites, it was necessary to adapt the FEB boxes to prevent the board from overheating. In GP300, a heat dissipation structure was added to the lid of the box.  In G@A, the top lids of the FEB boxes were replaced with a metal mesh. A mesh could not be used in GP300 because sandstorms in the Gobi desert would introduce sand into the electronics. In addition, G@A was fitted with a firmware controlled automatic hibernation based on the ADC chip temperature.


\subsection{Overall hardware system performance}
\label{sec:TF}

\begin{figure}[t!]
 \centering
 \includegraphics[width=.95\textwidth]{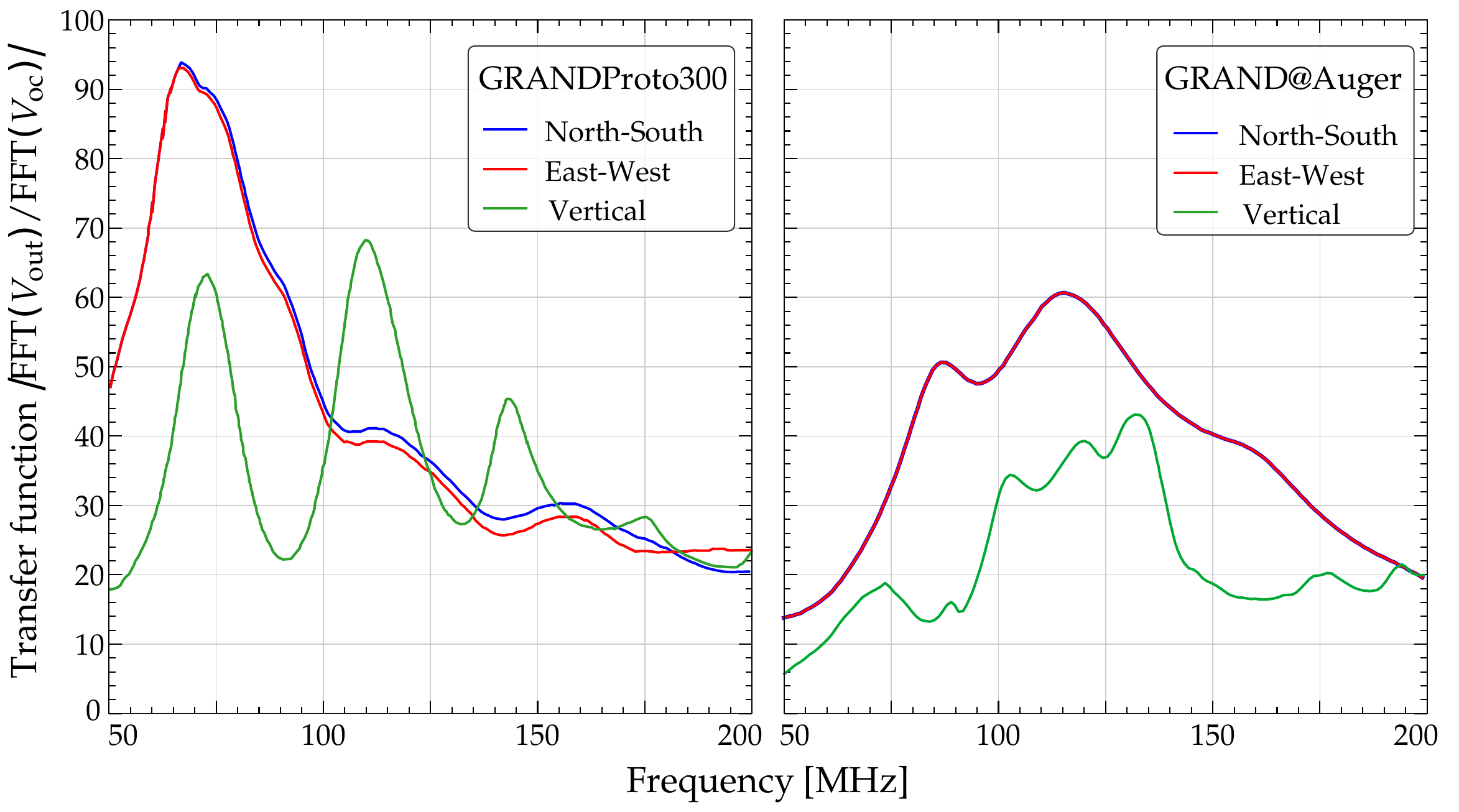}
 \caption{\textbf{Transfer function of a GRAND detection unit}, simulated using {\sc GRANDlib}~\cite{GRAND:2024atu}. \textit{Left:} In GP300, with a 20-dB gain at the FEB. (Differences between this figure and Fig.~8 in Ref.~\cite{GRAND:2024atu} are due to a revision of the S-parameters in the RF chain currently used by {\sc GRANDlib}.)  \textit{Right:} In G@A, with a 0-dB gain at the FEB. For G@A, the transfer function for the North-South polarization is exactly superimposed on the East-West polarization.}
 \label{fig:transferFuncs}
\end{figure}

The complete hardware system performance is simulated via the implementation of the Radio-Frequency chain (RF chain) in {\sc GRANDlib}~\cite{GRAND:2024atu}, a software tool developed and tailored specifically to the needs of the GRAND Collaboration. The RF chain receives as input an open-circuit voltage, $V_{\rm oc}$, which emulates the output voltage from the antenna arms. The final output voltage of {\sc GRANDlib}, $V_{\rm out}$, is the resulting signal after passing through the matching network, LNA, cables, connectors, and FEB. Reference~\cite{GRAND:2024atu} contains more details of {\sc GRANDlib}, the RF chain, and its implementation, which is modular and can be adapted to the different setups of GP300 and G@A. The RF chain is described via Scattering parameters (S-parameters), which quantify the reflection and transmission of the signals as they travel through the different components of a GRAND DU. The S-parameters of GP300 and G@A were measured separately in the laboratory and fed to {\sc GRANDlib}.

Figure \ref{fig:transferFuncs} shows the resulting simulated overall hardware system response as a function of frequency for GP300 and G@A.  The response is quantified through the transfer function, defined as the ratio $V_{\rm out}/V_{\rm oc}$, measured with respect to the base of the DU (see also Fig.~8 in Ref.~\cite{GRAND:2024atu}). Within GP300 and G@A, the transfer functions of the North-South and East-West polarizations are similar to each other, a feature also reflected in their VELs (Fig.~\ref{fig:ant_prop}, top). For both arrays, the system response is strongly frequency dependent. 

\subsection{Power supply and electromagnetic shielding}

In both arrays, the power supply of each DU consists of a solar panel, a charge controller, and a battery. The power consumption of a single DU is roughly 15\,W. The power supply components were specified to ensure that the DUs can operate for one week on a full battery, even without input from the solar panel. In GP300, the solar panels have a nominal power of 180\,W, with a maximum power voltage of 19.5\,V. Low-noise charge controllers (Genasun GV-10-Pb-12V) were chosen to efficiently charge the battery from the solar panel and provide power to the FEBs. Cold-resistant gel batteries were selected, operating at a nominal output voltage of 12\,V with a capacity of 200\,Ah. 

In G@A, the AERA electronics compartment and power harvesting were mostly re-used, as the power requirements of the prototype GRAND FEB match the AERA electronics quite closely. The solar panel was the same originally installed in the AERA-II stations (Kyocera 135~W), with a nominal power of 135\,W and maximum output voltage of 16\,V. New charge controllers, of the same model as deployed at GP300, were installed to replace the original AERA ones. The batteries were also exchanged for lead-acid batteries with a nominal output voltage of 12\,V and 180\,Ah capacity. Overall, the final power supply for G@A has a smaller capacity than for GP300. As a result, G@A DUs may need to hibernate during Argentinian winters, depending on their battery levels. 

In GP300 and G@A, the charge controllers are housed in Faraday cages. To further isolate them from noise and RFIs, a filter (Spectrum Control 1250-054) was added to the output of the charge controller. The Bullet is also housed in the FEB box at G@A, and, due to the heat-dissipation modifications, in a separate Faraday cage at GP300. 


\subsection{Communications}

Communication between the local and central DAQ softwares is done over TCP/IP sockets for both G@A and GP300. Upon request from the central DAQ, the complete event is sent over the wireless network, which is created through M5 (G@A) or AirMax (GP300) \texttrademark{Ubiquiti} Bullets and Rockets. The rocket is placed at the Central Station and each GRAND DU has a bullet connected to its FEB by an ethernet cable. The parabolic communication antenna connects to the bullet and focuses its transmission to the rocket position. The rocket is connected to a planar antenna aimed to receive signals from multiple locations. A proprietary time sharing protocol is used to enable fast and stable communication. For G@A, the bullets and rocket were provided by the Pierre Auger Collaboration. 

Bi-directional transmission tests were performed point-to-point, {\it i.e.}, single DU to DAQ, in both arrays, using a fixed file size. The uplink speed, from DU to DAQ, was on average 2.9~MB~s$^{-1}$ for GP300 and 2.1~MB~s$^{-1}$ for G@A. The downlink speed, from DAQ to DU, was on average 4.2~MB~s$^{-1}$ for GP300 and 1.1~MB~s$^{-1}$ for G@A. 

For the overall network to the central DAQ computer, the Ubiquity system achieves a transfer rate of 9~MB~s$^{-1}$, which is shared between all DUs. Taking into account typical GRAND event sizes, a bandwidth of 2 KB~s$^{-1}$ per station is needed. Therefore, communication between the DUs and the central DAQ does not limit the total throughput of the system. However, data transfer across ethernet has some latency, so we allow up to 3 seconds to ensure that all data is received by the central DAQ. Central trigger decision making---see Section \ref{sec:trigger_DAQ} for details---and transfer back can take up to an additional second, which results in a total time of 4 seconds for which data needs to be buffered locally. Thus, the transfer from FGPA to DDR memory in the DU FEB limits the total transfer rate to roughly 1400 Hz. 


\section{Data acquisition processes}
\label{sec:trigger_DAQ}

The main science case of GRAND---the study of ultra-high-energy particles~\cite{GRAND:2018iaj}---involves the radio-detection of extensive air showers. These emit electromagnetic radiation in the radio frequency range, which induces a temporary distortion of the local electric field with a duration of several nanoseconds. The radio-frequency pulse generates a high-amplitude (relative to the average baseline), short-duration oscillation at the GRAND DU ADC output, which allows for its detection. However, Radio Frequency Interference (RFI) pulses, present in the detector array region, can have similar characteristics. The main differences to air shower-induced pulses are larger low frequency contributions and/or arriving in a train of pulses. Thus, air shower experiments that rely solely on their radio-detection, like the GRAND prototype arrays, must have a data acquisition chain with a self-triggering procedure that efficiently selects detected events which most resemble those expected from air showers.

\subsection{Firmware and self-trigger algorithm}

\begin{figure}
    \centering
    \includegraphics[width=.65\linewidth,trim={.2cm .4cm .4cm .3cm}, clip]{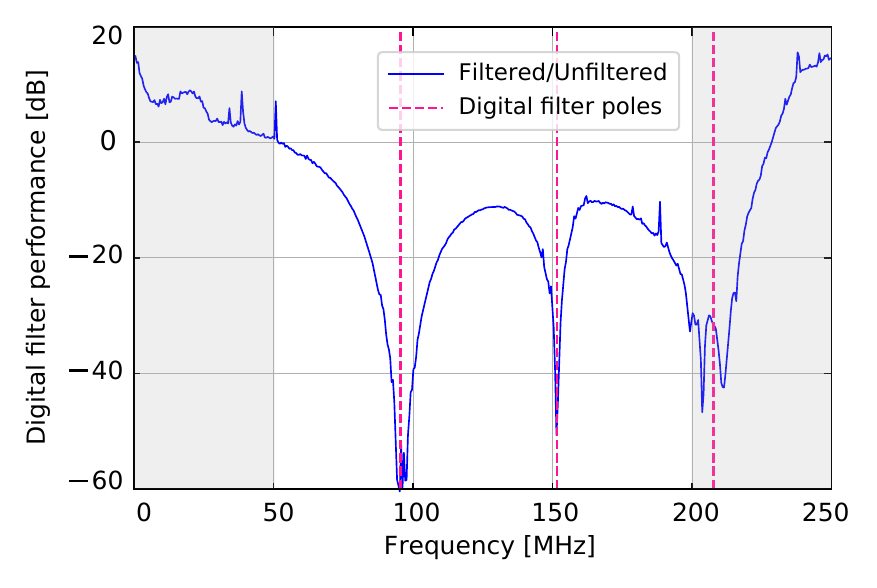}
    \caption{\textbf{Measured performance of the digital notch filters on the prototype GRAND FEB.} The filter pole frequencies were set at 95.5, 151.5, and 207.7~MHz (in pink), and were applied to  North-South polarization data. We show the ratio of the filtered and the unfiltered data (in blue), in which the decrease in intensity at the pole frequencies demonstrates proper filtering performance.}
    \label{fig:filter_performance}
\end{figure}

The lowest level of data handling is carried out by the firmware running on the FPGA and CPU chip---specifications are given in Section \ref{sec:FEB}. The firmware continuously retrieves the digital data from the ADC. It is configured such that, for each input sample, the previous 32 (64 ns) are used to evaluate the true baseline and perform a dynamic baseline subtraction. Sixteen independent second-order digital Infinite Impulse Response (IIR) notch filters \cite{960414IIRfilter}, four per signal input, can be used to digitally shape the signals. Each filter has a pole frequency and width that can be individually configured. They allow us to first remove single-frequency background and then feed the resulting time-traces to the trigger algorithm. Figure \ref{fig:filter_performance} shows the performance of the digital filter using three independent notch filters on data collected in the G@A prototype array. If signal shaping is selected, the shaped signals are used for self-triggering. Otherwise, raw data are used. For the final readout, it is possible to select raw or shaped data.


\begin{figure}[t]
 \centering
 \includegraphics[width=0.8\textwidth]{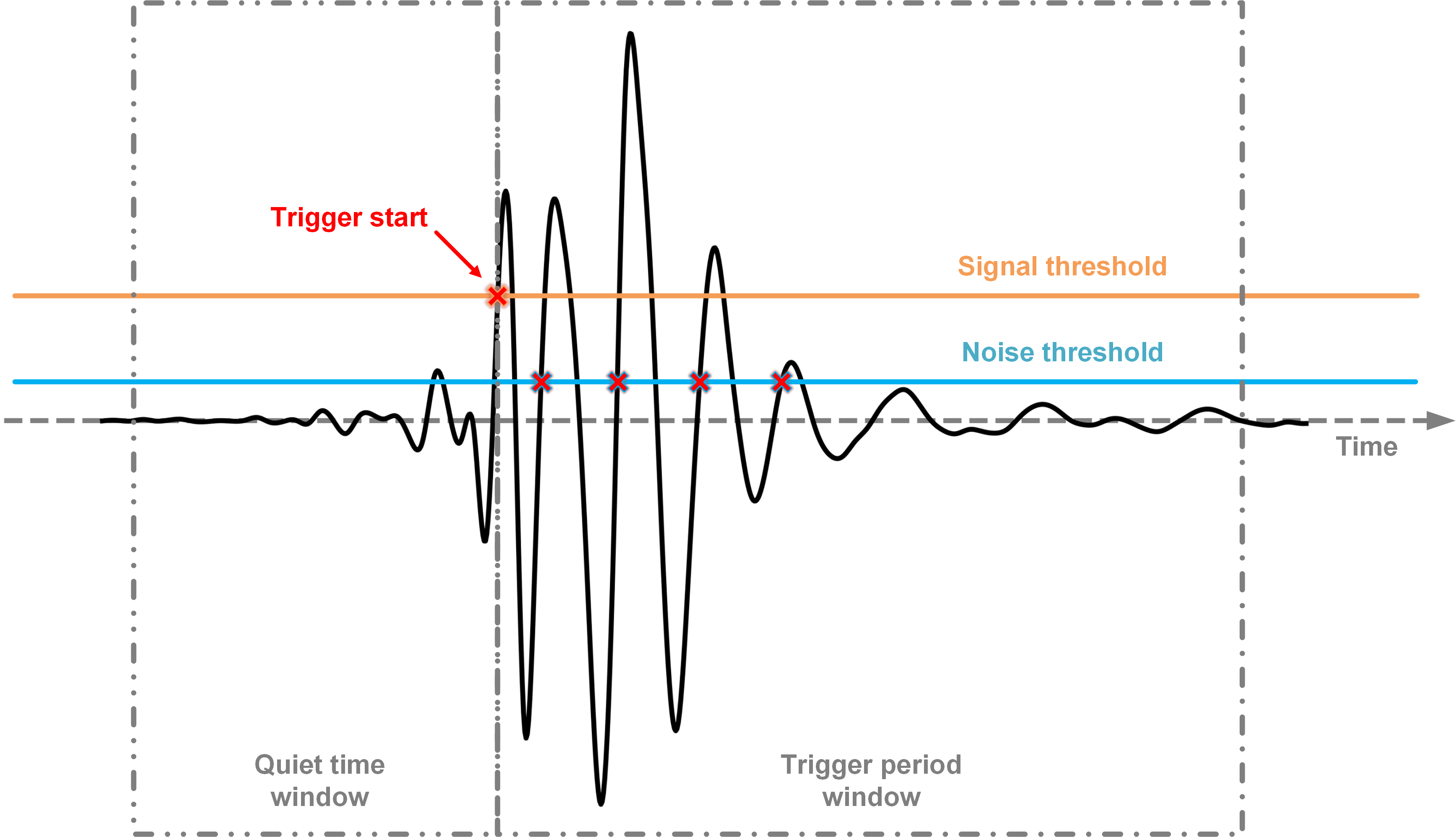}
 \caption{\textbf{Local Trigger flow for an incoming signal at a GRAND detection unit.} The signal depicted represents a typical air shower-induced electric field pulse in a limited frequency band. The configurable trigger parameters are indicated w.r.t. the radio-frequency pulse, with time increasing to right.}
 \label{fig:trigFlow}
\end{figure}

The firmware performs the Local Trigger, at the DU level. The aim is to select air-shower candidates while rejecting background RFI pulses by efficiently identifying peaks using the difference of two consecutive ADC samples. Figure \ref{fig:trigFlow} shows a representative flowchart of the trigger algorithm. It starts when the amplitude of the input signal exceeds the preset ``signal threshold", \textit{i.e.}, when a signal threshold crossing occurs. This condition is valid as long as there was no such crossing in the preceding quiet time window. In the time period following the signal crossing, called the ``trigger period window'', the algorithm counts the number of crossings of a lower ``noise threshold".  If the number of noise-threshold crossings is between a predefined allowed minimum and maximum values, and the time interval between any two adjacent noise-threshold crossings does not exceed a predefined maximum value, the trigger process is completed and the signal is accepted.  

The width of the time windows, the signal and noise threshold levels, the allowed maximum number of crossings of the noise threshold, the maximum time between successive crossings, and the time-series length are adjustable parameters within the firmware. They are set by the Central DAQ prior to each data acquisition run. The values are chosen to best reject long signals and/or pulse trains and to prevent digital glitches from corrupting the voltage traces. This trigger flexibility allows us to adapt to the different noise and background environments of each array. 

Although many trigger configurations were tested during the commissioning period of both prototype arrays, there was a common convergence. The signal threshold value is usually set to five times the standard deviation of the typical time-series noise level. The noise threshold is configured to be on the order of two times the amplitude of the same noise level. In GP300, keeping these threshold levels fixed is enough to ensure a reasonable trigger rate. Due to the more frequent narrow and broadband RFIs in G@A, these were used as initial threshold levels, which are dynamically adjusted by the firmware to maintain the trigger rate up to a certain value.

The configured quiet time window is of the order of one $\mu$s, followed by a trigger period window of half a $\mu$s. Within this window, most often less than five noise threshold crossings are allowed, with a maximum time between crossings of the order of 50~ns. The final recorded time-series have lengths of either 2048 or 4096~ns, depending on the data acquisition requirements for frequency resolution. Given the trigger conditions, the detector deadtime is at least 1.5~$\mu$s. 

When the trigger conditions are met, the different information fractions that make up the event are placed in the appropriate registers. For each time-trace, two registers are used---pre and post trigger. One register is filled with the event header and the second with the GPS information. The firmware indicates when all these processes are completed, and the Local DAQ software, running under PetaLinux, initiates the transfer from the registers into memory.  

Two firmware versions were tested, one on each prototype array, which differed in the method of data transfer into memory. The firmware version deployed in GP300 starts by reading event packages by the Advanced eXtensible Interface (AXI) bus, as long as the FIFO queue for storing event data is not empty. Given the quieter electromagnetic environment of GP300, the AXI speed is enough to deal with the trigger rate. However, the higher trigger rate at the G@A site required a faster memory transfer method. Therefore, the software uses Direct Memory Access (DMA transfer), moving all parts of each triggered event into the DDR memory. Hence, each fragment is moved to the correct memory position and the final event is created. The event timestamp, corresponding to the time at the end of the trigger period window, is recorded and sent to the Central DAQ software. The complete events are locally stored in the ring buffer until requested by the Central DAQ or overwritten.

Next to recording air showers, the arrays also keep track of the electromagnetic environment conditions via unbiased triggering. This corresponds to periodically recorded data, independent of the measured voltage levels. The unbiased data are suitable for monitoring the local RFI background and for calibrating the DUs by using the periodic variation of the Galactic synchrotron emission---further details are provided in Section \ref{sec:spectra}. Unbiased triggering, by default, takes place once every 10~seconds (a pre-defined 20-Hz mode is also available). In the G@A firmware version, it is possible to set the unbiased triggering frequency between 1\,Hz and 100\,kHz. GP300 and G@A perform both transient and unbiased triggering.

Trigger efficiency as a function of input rate was measured with the GP300 firmware using a signal source that generated four periods of a 150-MHz sine wave fed into one channel of the GRAND FEB. A 100\% trigger efficiency was measured up to a repetition rate of 1434~Hz, a value higher than the nominal 1~kHz trigger rate at DU level. These results also apply to G@A, which has registered trigger rates of about 1~kHz on-site without losing functionality. Under typical conditions, the DU trigger rate is on the order of 100~MHz for both prototype arrays. An efficient self-trigger algorithm is a main priority to achieve the GRAND science case and will be further detailed in a distinct publication.

\subsection{Data acquisition software}

\begin{figure*}[t!]
  \centering
  \includegraphics[height=7cm]{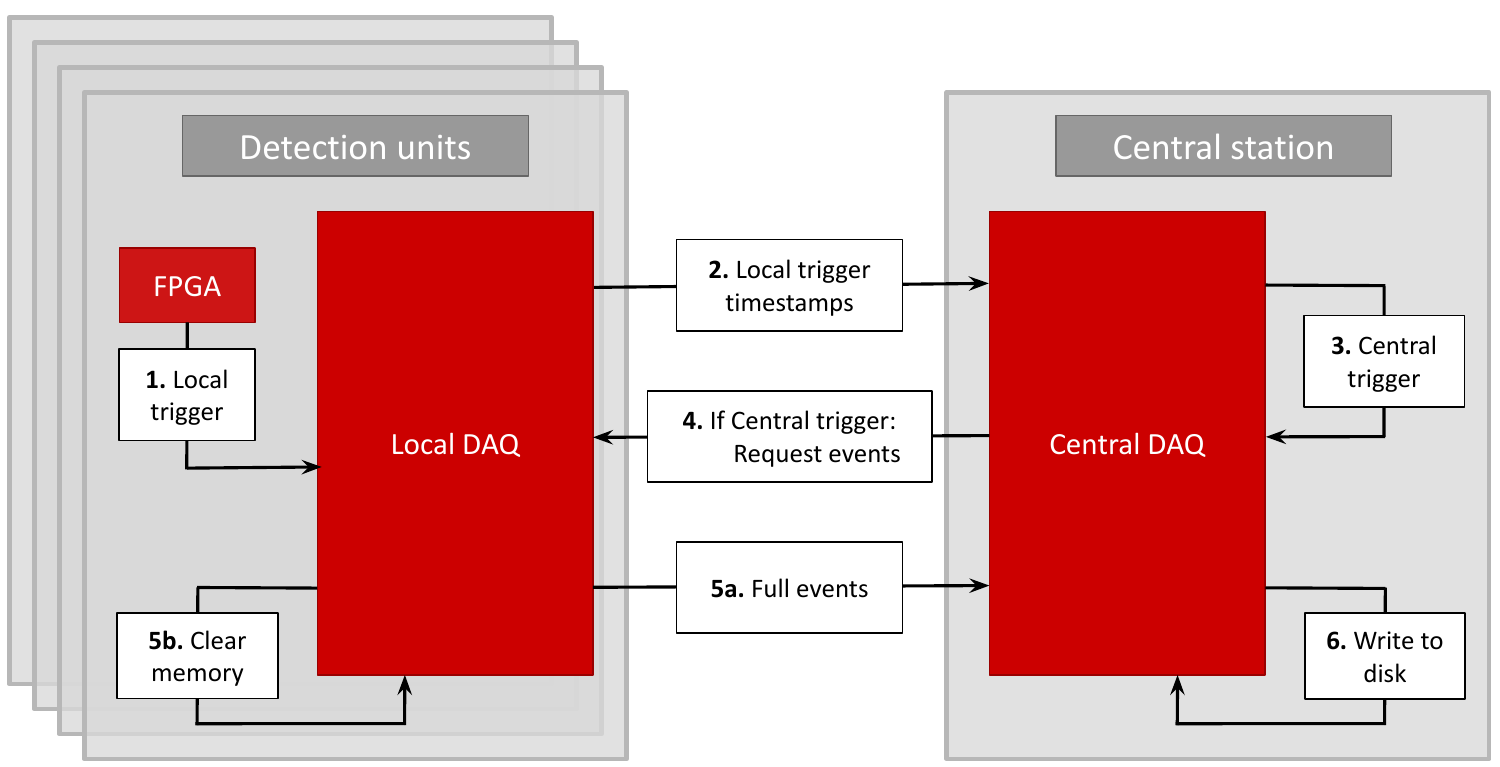}
  \caption{\textbf{Data acquisition flow for triggered events in the GRANDProto300 and GRAND@Auger.} The cycle repeats indefinitely while the data-taking is ongoing. Unbiased triggered events bypass the Local Trigger at the FPGA but still go through the same DAQ pipeline.}
  \label{fig:soft3}
\end{figure*}

Figure~\ref{fig:soft3} shows a diagram of the GRAND DAQ software. It operates on two levels---the Local DAQ at each DU and the Central DAQ at the Central Station Computer of each array---which interact to manage the complete data flow. The Collaboration developed two separate versions of the DAQ software for GP300 and G@A to assess different implementation strategies. For the next stages of GRAND, a unified and optimized version will be developed.

The Local DAQ temporarily saves both types of triggered data, transient and unbiased, in local storage, used as a circular buffer, step (1) in Fig. \ref{fig:soft3}. The GP300 firmware uses the Petalinux 30~MB memory, while the G@A firmware uses a 512\,MB DDR memory. Events are stored locally, while the Central Trigger, running on the Central DAQ, determines if they are to be retrieved and saved permanently. 

The Central Trigger decision is based on the timing and transient trigger condition information that the Local DAQ sends to the Central DAQ, step (2) in Fig. \ref{fig:soft3}. The trigger decision, step (3), is determined by configurable parameters, such as the maximum time difference between the measured signals at different DUs and the number of DUs that detect a signal in coincidence within this time window. As soon as there is a positive decision by the Central Trigger, the Central DAQ requests the full event information from the Local DAQ, step (4) in Fig. \ref{fig:soft3}. The full event information is sent over the wireless network to the Central DAQ and its event-building process, step (5a), and the Local DAQ memory is cleared, step (5b) in Fig. \ref{fig:soft3}. This process combines the requested full event information and the local trigger timestamps into a single event and writes it to the corresponding data file on the local disk of the Central DAQ, step (6) in Fig. \ref{fig:soft3}. 

All coincident transient triggered data that satisfy the Central Trigger are recorded in Coincident Data (CD) binary files. Furthermore, a random selection of local transient triggered events can be promoted to a Central Trigger and be saved as well. These events have no coincidences with other DUs and are used to monitor the trigger algorithm and the noise environment. They are stored separately in Unit Data (UD) binary files. Unbiased triggers are always promoted to Central triggers and are saved in Monitoring Data (MD) files.

\section{Data management}\label{sec:data_management}
\subsection{Storage}

The final data files of each trigger type---CD, UD, and MD---are saved in the local storage of the Central DAQ computer on GP300 and G@A. For GP300, the binary files are sent in real time to a computer center in Nanjing, China, through a wireless bridge. The files are then manually moved to the CC-IN2P3 cluster in Lyon, France. For G@A, new files on disk are copied daily to the central storage in CC-IN2P3 using the 4G network. This copying does not interfere with the data-taking or the bandwidth used by the Pierre Auger Collaboration. In case of high RFI conditions or failed off-site data transfer, the DAQ computer has enough disk space to store several months of triggered data.

Automatic copying requires a local bash script to run automatically ({\it i.e.}, a \texttt{cron} job) on the central computer for G@A and semi-manually for GP300. This script first searches for files not already transferred to CC-IN2P3 and stores this list in a simple local SQLite database. This database is used to keep track of previously transferred files and to determine if a file needs to be sent. Afterwards, the new data files are relayed to the corresponding directories on CC-IN2P3 using \texttt{rsync} and then flagged into the local SQLite database. 

When data dispatch is completed, a second script is triggered in CC-IN2P3 that registers all newly transferred files in the main database and launches the conversion into the GRANDRoot format, described in Ref.~\cite{GRAND:2024atu}. The GRANDRoot files are moved to the corresponding directories and registered in the main database. 


\subsection{Quality monitoring}
\label{sec:data_monitoring}

To ensure adequate data quality during the deployment and commissioning phases of GP300 and G@A, a few key parameters are tracked in every DU. A complete event consists of the triggered ADC time-traces and the slow control parameters, such as the battery level at the board input, the GPS information, and, in G@A alone, the ADC and FPGA chip temperatures and the trigger rate. 

Through the unbiased triggered data, these parameters can be continuously monitored. Their behavior was used not only to commission the prototype arrays but also to validate the firmware and the local and central DAQ software. Other features---such as their Root Mean Square (RMS), Fast Fourier Transforms (FFTs), and Power Spectral Densities (PSDs)---are extracted from the ADC traces in the time domain to monitor day-night effects and detect local RFI transients. In particular, the PSDs can be used to characterize the environment radio-frequency background, the thermal noise from the electronics, and the Galactic noise. These parameters are checked daily to ensure optimal operation of the prototypes. We monitor their behavior over time, comparing them between DUs to achieve consistent data quality across the array.

The commissioning is facilitated by a data monitoring web page.
The page is updated daily at 05:00 CET, by a \texttt{crontab} job that runs a series of scripts at CC-IN2P3 on the data from both arrays. The DU uptime, the mean RMS of the last 24 hours of traces, battery levels, and temperatures are tracked as a function of time. The averaged PSDs are also displayed for easy comparison between channels and between DUs. The page also stores the averages over the last 7 days and the last 30 days. There is flexibility to present the data of all DUs together or of only the desired ones.


\section{Background and noise characterization}
\label{sec:spectra}

\begin{figure*}[t!]
  \centering
  \includegraphics[width=\textwidth]{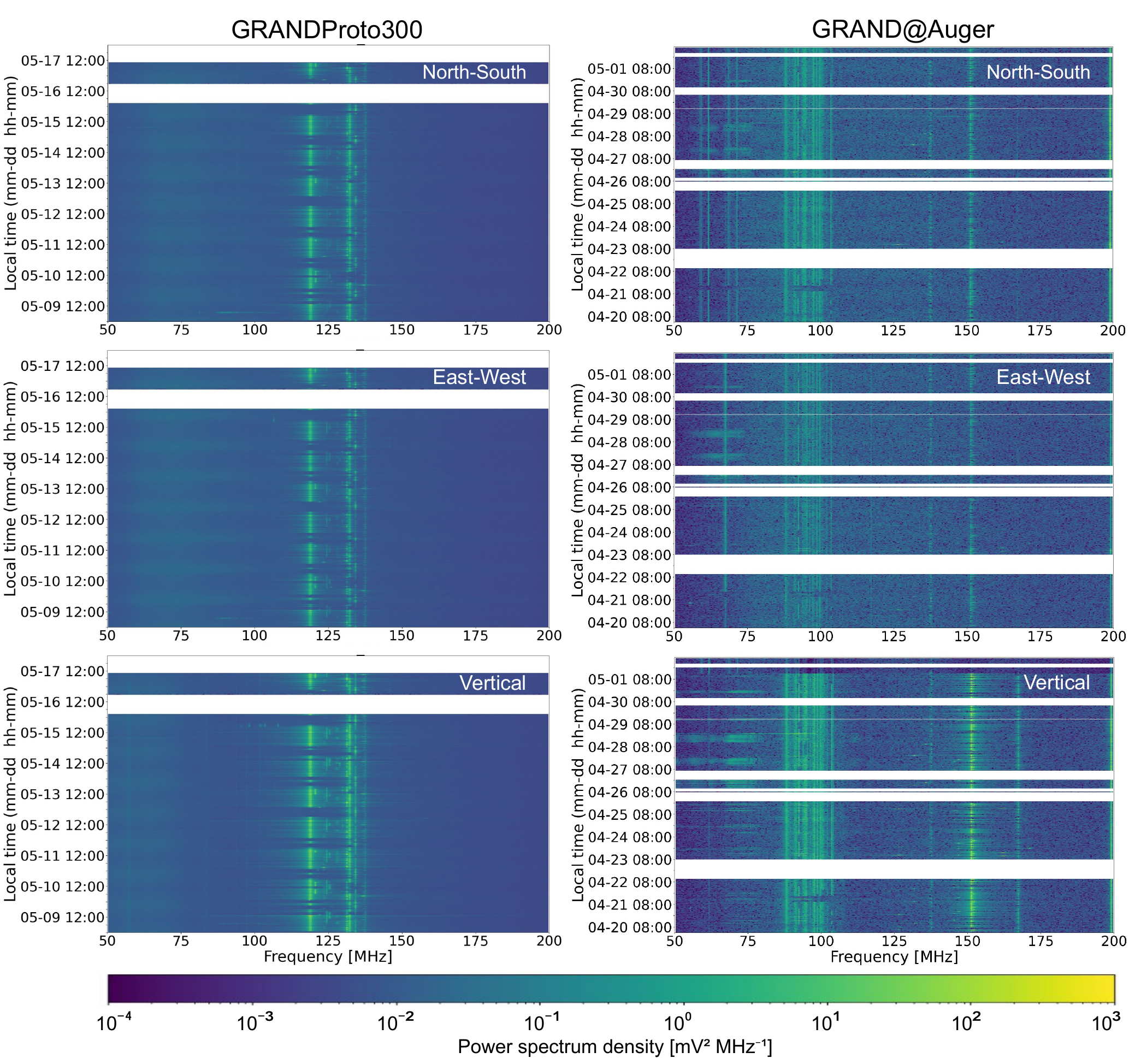}
  \caption{\textbf{Spectrograms measured by GRAND detection units.}    Each row shows spectrograms for a different antenna polarization: North-South (\textit{top}), East-West (\textit{center}), and vertical (\textit{bottom}). The spectrograms are built from voltage time-series with a duration of 2~$\mu$s. The frequencies are binned in intervals of 0.5~MHz and the local time is binned in intervals of 30~minutes. White horizontal lines correspond to periods of hibernation shutdown. In both DUs, vertical lines of high power spectrum density correspond to constant radio background. \textit{Left:} Spectrograms for DU 6 at GP300, using data collected over nine days in May of 2025. \textit{Right:} Spectrograms for DU 83 at G@A, using data collected over twelve days in April of 2024.}
  \label{fig:spectrograms}
\end{figure*}

GP300 and G@A have operated for about two years since the end of their deployment (with GP300 changing from comprising 13 to 65 DUs). Throughout this period, enough unbiased triggered data was collected to characterize the performance of the DUs and the local electromagnetic environment. These studies were performed in the frequency domain, where the complete antenna readout chain is better understood and background sources can be identified.
In addition, as the DUs are deployed relatively close together, working in the frequency domain allowed us to verify that all of them showed similar response, which is needed to understand the relative calibration between DUs.

Figure~\ref{fig:spectrograms} shows spectrograms generated using unbiased triggered data collected by two DUs, one from GP300 and one from G@A, over a period of about nine and twelve days, respectively. These spectrograms were calculated from voltage time-series measured at the ADC input and are representative of the standard behavior of all DUs in their respective arrays. For GP300, the PSDs are largely free from strong narrowband RFIs in all polarizations, except within 118--140 MHz, where anthropogenic background is identified. In addition, Figure~\ref{fig:spectrograms}, left, shows an apparent daily modulation in the PSD.

For G@A, several broad and narrowband RFIs are present in the three polarizations, some continuous throughout the data acquisition time and some not. Of the intermittent contributions, one at around 70~MHz can be seen in all polarizations, where there is an increase of amplitude for a few hours with no clear periodicity. In the North-South polarization, the four lines within 60--80 MHz correspond to the known emission of the AERA beacon, set up at the CRS (see Fig.~\ref{fig:arrays}). The higher variability of background and noise at G@A compared to GP300 is reflected in the larger granularity of the spectrograms.

In principle, unbiased data are collected with a 100\% duty cycle, unless the acquisition is interrupted. For instance, in Fig.~\ref{fig:spectrograms}, the white horizontal lines correspond to periods of DU hibernation, in which data acquisition was suspended and then resumed autonomously once conditions are appropriate. The G@A DU had more hibernation periods than the GP300 DU, a reflection of the harsher environment in Argentina, leading to more periods with insufficient sunlight to power the batteries and more instances of overheating. In general, the electromagnetic environment at GP300 is considerably quieter than at G@A, but there are RFI-free frequency bands at both sites. Nevertheless, the background in several frequency ranges is substantial and can negatively influence the self-trigger capabilities of the system, especially in G@A.

\begin{figure*}[t]
  \centering
  \includegraphics[width=.85\textwidth,trim={.7cm 1.2cm 2.cm 2.7cm}, clip]{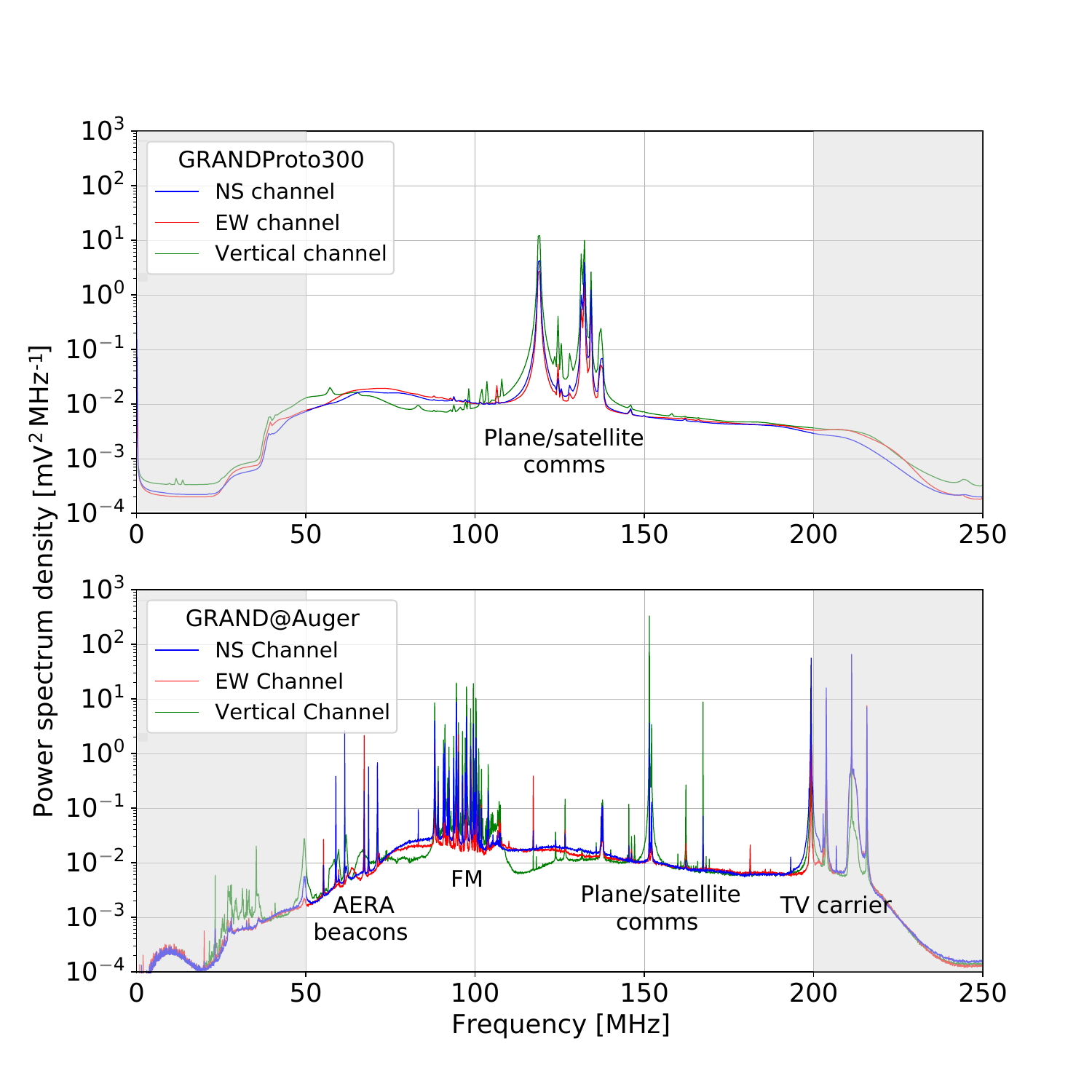}
  \caption{\textbf{Time-averaged power spectrum density measured by GRAND detection units.} A few prominent peak sources are highlighted. \textit{Top:} Spectra for GP300, averaged over one day of May-2025 data of DU 6. The average spectrum was obtained from voltage time-series with a duration of 2~$\mu$s and a frequency resolution of 0.5~MHz. 
  \textit{Bottom:} Spectra for G@A, averaged over 8 hours, from 2 to 11~a.m. in local time, of March-2024 data of DU 83. The average spectrum was obtained from voltage time-series with a duration of 16~$\mu$s and a frequency resolution of 62.5~kHz, collected in a dedicated run. }
  \label{fig:spectra}
\end{figure*}

Figure~\ref{fig:spectra} shows the time-averaged PSDs of GP300 and G@A, where the background RFI contributions can be seen more clearly. Similarly to the spectrograms, unbiasedly triggered voltage traces, also measured at the ADC output, were used to evaluate the average PSDs. 

\begin{table}[b!]
    \centering
    \begin{tabular}{ccc}
        Frequency [MHz]  & Channel & Source \\ \hline
        87—108 & NS, EW, Vert  & FM band \\
        108--117.975 & NS, EW, Vert  & Aeronautical radionavigation \\
        117.975--137 & NS, EW, Vert  & Aeronautical mobile \\
        137--138 & NS, EW, Vert  & Space operations and research, \\
        & & satellite meteorology \& mobile comms. \\
    \end{tabular}
    \caption{\textbf{Peaks in the GRANDProto300 power spectrum density and their sources.} The peaks are shown graphically in Fig.~\ref{fig:spectra}.}
    \label{tab:fft_gp300}
\end{table}

\begin{table}[b!]
    \centering
    \begin{tabular}{ccc}
        Frequency [MHz]  & Channel & Source \\ \hline
        55.2 & NS, EW, Vert  & TV channel 2, video\\
        58.9 & NS & AERA beacon \\
        59.8 & NS, EW, Vert & TV channel 2, audio \\
        61.5 & NS, EW, Vert & AERA beacon \\ 
        67.3 & NS, EW, Vert & TV channel 4, video\\
        68.5 & NS & AERA beacon \\
        71.2 & NS, Vert & AERA beacon \\
        83.2 & NS & TV channel 6, video \\
        88--104 & NS, EW, Vert & FM band \\
        &  & 18 radio stations \\
        117.2 & EW & Aviation \\
        147.2 & Vert & Satellite comms. \\
        151.5 & NS, EW, Vert & Satellite comms. \\
        152.2 & NS, EW, Vert & Satellite comms. \\
        167.3 & NS, EW, Vert & $\cdots$ \\
        199.3 & NS, EW, Vert & TV channel 11, video \\
        202.8 & NS, EW, Vert & $\cdots$\\
        203.8 & NS, EW, Vert & TV channel 11, audio \\
        206.7 & NS & $\cdots$ \\
        211.2 & NS, EW, Vert & TV channel 13, video \\
        215.7 & NS, EW, Vert & TV channel 13, audio \\
    \end{tabular}
    \caption{\textbf{Peaks in the GRAND@Auger power spectrum density and their sources.} The peaks are shown graphically in Fig.~\ref{fig:spectra}.}
    \label{tab:fft_gata}
\end{table}

For GP300, the PSD is mostly free of peaks, except in the 118--140~MHz range where there are strong contributions from plane and satellite communications. These sources have high intensity and are spread over multiple frequency bins, collectively defining a frequency band that will be excluded in analyses. Above 140~MHz, there is a slight drop in the PSD baseline level. Since this range is dominated by electronic noise, only gain effects are expected, which is consistent with the drop seen in the system performance shown in the transfer functions in Fig.~\ref{fig:transferFuncs}. For G@A, we performed a dedicated run with longer time-series and finer frequency resolution (see the caption of Fig.~\ref{fig:spectra}) to resolve individual background RFI sources more clearly. In Fig.~\ref{fig:spectra}, the drop in the PSD at low frequencies is due to the impedance mismatch between the LNA and the antenna.

Tables \ref{tab:fft_gp300} and \ref{tab:fft_gata} show the identified sources of several of the peaks of the GP300 and G@A spectra in Fig.~\ref{fig:spectra}. In particular, the G@A PSD characterizes the electromagnetic background at the Pierre Auger site at frequencies much higher than either AERA or the Pierre Auger Radio Detector are capable of measuring. This can aid and complement analyses of the Pierre Auger Collaboration---an example of the synergy between the Pierre Auger Observatory and GRAND.


\begin{figure}
    \centering
    \includegraphics[width=\linewidth,trim={2.9cm .5cm 3.8cm 1cm}, clip]{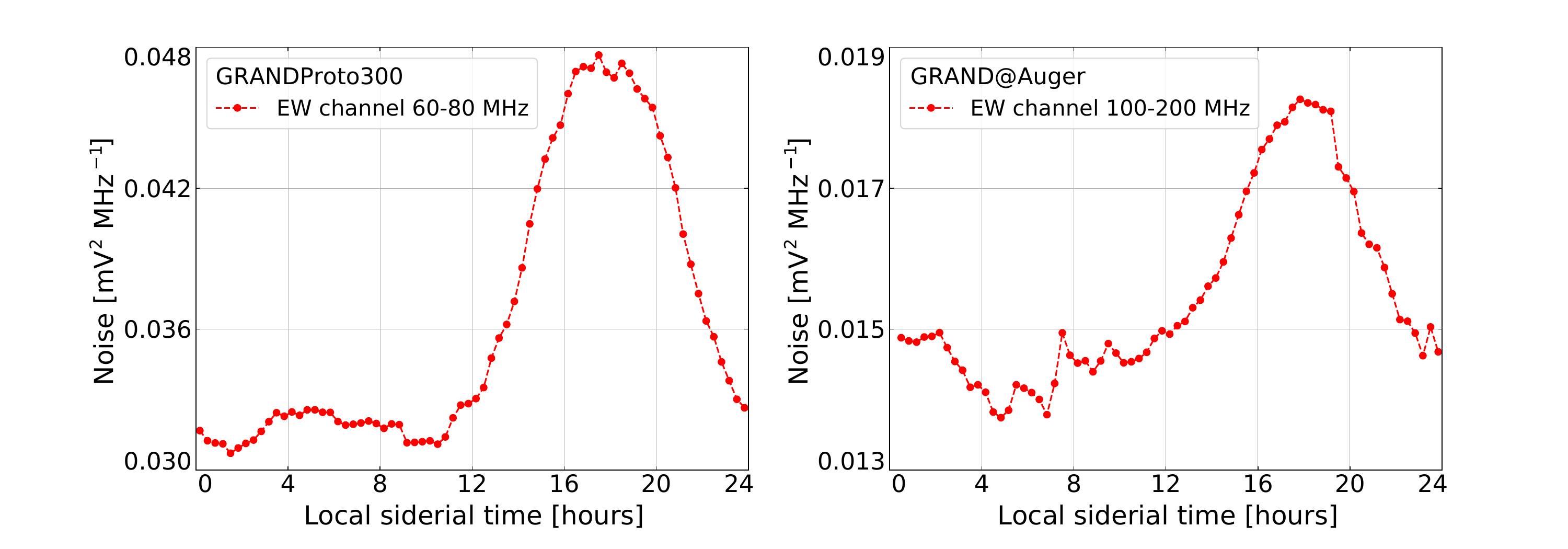}
    \caption{\textbf{Galactic synchrotron emission measurement with GRAND detection units.} Power Spectrum Densities (PSDs), built from the East-West polarization, were averaged over a period of two months. \textit{Left:} For GP300, unbiased data from June and July 2025 collected by DU 6 were used, in the frequency range of 60--80~MHz. \textit{Right:} For G@A, unbiased data from DU 83 were also used in the frequency range of 100--200~MHz. The acquisition period was from March to June 2024, which, after a quality selection, summed up to 50 days.    A amplitude maximum around the expected peak Galactic emission is clearly visible for both prototypes.}
    \label{fig:gb}
\end{figure}


Identifying the frequency of narrowband RFIs facilitates their exclusion from the measured time-traces. In this way, we can obtain large RFI-free frequency bands, despite the high background content at both sites. This allows us to perform a first characterization of the stochastic noise level at each prototype array. The two main contributions expected within the GRAND working frequency range are thermal noise from the electronics and the Galactic synchrotron emission. The latter originates mainly from the Galactic plane, yet it is brightest in the inner Galaxy region. Thus, the emission reaches peak amplitude when the detector field of view is turned towards the Galactic center. If the GRAND DU thermal noise does not dominate the measurements, we expect to observe an increase in the average PSD between 16 and 20 hours in Local Sidereal Time (LST) corresponding to the Galactic emission.

Figure \ref{fig:gb} shows the average PSDs for limited frequency ranges---60--80~MHz for GP300 (left) and 100--200~MHz for G@A (right)---. For both setups, the data was averaged over a period of about two months, sufficient for a significant observation of the Galactic noise. The differences in frequency range are a reflection of the different optimization frequencies of the LNA design in each prototype array---the motivation for this choice is discussed in Section \ref{sec:setUp_antenna_lna}. Due to the frequency dependence of the Galactic emission, as well as the antenna's side lobes, it is expected that the absolute amplitude of the Galactic noise will be distinct for different frequency ranges. This is seen in the noise ranges of each plot in Figure \ref{fig:gb}, which are not the same but similar in magnitude. Nonetheless, we can state that both GP300 and G@A have detected the Galactic synchrotron emission. More detailed analyses are described in the references \cite{Ma:2025xgd, deErrico:2025qsa}.


\section{Summary and perspectives}
\label{sec:ending}

The Giant Radio Array for Neutrino Detection (GRAND) is a planned observatory of ultra-high-energy particles of cosmic origin: cosmic rays, neutrinos, and gamma rays.  GRAND aims to detect the electromagnetic emission from air showers initiated by their interaction in the atmosphere using large ground arrays of antennas. The next large deployment goal of GRAND is to have two arrays consisting of 10,000 antennas each in the Northern and Southern Hemispheres. To achieve this, the GRAND Collaboration has adopted a staged approach to deployment, with three ongoing prototype arrays to test and validate the detection and reconstruction methods. 

Three prototype arrays were deployed in 2023. GRAND@Nançay, in France, is a 4-antenna array used as a testbench for the Collaboration's European laboratories. GRANDProto300, in China, has gone through two stages of deployment, from an initial 13-antenna stage to its current 65-antenna stage. These stages act as pathfinders for the future final 300-antenna array. GRAND@Auger, in Argentina, stems from an agreement between the GRAND and the Pierre Auger Collaborations. The 10-antenna array was deployed on modified AERA stations. In this paper, we focus on the hardware, software, and early operation of GRANDProto300 and GRAND@Auger.

To adapt to local conditions and explore optimal designs for the final GRAND setup, the prototype array at each site featured a few different approaches to hardware and software. Throughout 2023, GRANDProto300 and GRAND@Auger were commissioned using daily data monitoring. The arrays have remained largely operational for about two years. Early data demonstrate that both systems can trigger on weak radio-frequency signals and are able to handle trigger rates up to 1 kHz. We provide a first characterization of the electromagnetic background and noise present at each site. The Galactic synchrotron emission is measured at both prototypes and will be used to refine the calibration of the GRAND detection units. In GP300, there are fewer RFI sources compared to G@A. Nonetheless, the configuration at each site is such that baseline levels of the PSDs are the same for both prototype arrays, around $10^{-2}$~mV$^2$~MHz$^{-1}$. 

The GRAND Collaboration aims to achieve its goals while keeping its carbon footprint low~\cite{Aujoux:2021kub}. Following the life cycle analysis of the current prototype GRAND DUs~\cite{carbon_footprint_GRAND}, we have identified two design modifications that will decrease the environmental impact of the experiment. First, changing the alloy composition of our mechanical structure's stainless steel~\cite{carbon_footprint_GRAND} would lower the impact by 66\%. Secondly, using longer duration batteries would diminish the impact almost linearly. These design changes will be integrated into the preparation for the production of the next batch of antennas that will be deployed to complete the GRANDProto300 array.

Future stages of the GRAND experiment will face diverse environmental and electromagnetic background and noise conditions, due to the multiple geographical locations. The demonstrated adaptability of the GRAND hardware and software indicates the feasibility of combining data from the different 
arrays. This is essential for achieving full-sky coverage and large-scale exposure of future stages of GRAND. Presently, the data collected by GRANDProto300 and GRAND@Auger are being searched for air showers initiated by ultra-high-energy cosmic rays. In GRANDProto300, we have identified the first cosmic-ray candidates~\cite{Lavoisier:2025ase, Guelfand:2025hqr, Ferriere:2025csu}.  In GRAND@Auger, the detection of cosmic-ray candidates is validated through their coincident detection by the Pierre Auger Observatory and a first candidate has been observed~\cite{deErrico:2025qsa}.


\section*{Acknowledgments}

\noindent
The GRAND Collaboration acknowledges the significant contributions to this manuscript of Beatriz de Errico, for her work on data management and analysis, site survey and construction, DAQ software debugging, manuscript writing, and figure preparation; and of Shen Wang, for his work on GP300 site survey and construction, hardware verification, and manuscript planning and writing.  In addition, the Collaboration acknowledges significant contributions from the following Collaboration members: Mauricio Bustamante (coordination, manuscript writing, figures), Nicoleta Cucu Laurenciu (firmware development and trigger algorithm implementation), Bohao Duan (construction, GP300 DAQ software development), François Legrand (section on data storage), Pengxiong Ma (construction, RFI identification, GP300 data management and analysis), Frédéric Magnard (communications at GP300, DAQ and communications at Auger), Stavros Nonis (data analysis, RF chain implementation, and figures), Lech Piotrowski (GP300 DAQ software), Daniel Szálas Motesiczky (LNA design, electronics validation), Xishui Tian (laboratory trigger tests), Anne Timmermans (data monitoring web page), Charles Timmermans (site survey, construction, DAQ software development and debugging, data management and analysis, manuscript writing and revising, figures), Feng Wei (LNA design), Xin Xu (antenna response, RF chain optimization, data analysis), Xing Xu (construction, GP300 hardware verification, firmware development and trigger algorithm implementation), Pengfei Zhang (site survey, redesign and functional test of detection unit, construction, data analysis, RF chain generation), Yi Zhang (site survey and construction, data analysis, manuscript writing).  The GRAND Collaboration is grateful to Arno Engels (Radboud University) for the mechanical design of the detection units, to Fei Gao (Xidian University) for that and for the thermal design of the detection units, to the local government of Dunhuang during site survey and deployment approval, to Yu Tang for his help on-site at the GRANDProto300 site, and to the Pierre Auger Collaboration, including  the staff in Malarg\"ue, for the warm welcome and continuing support, and the Auger Publications Committee, especially Silvia Mollerach, for their feedback.
The GRAND Collaboration acknowledges the support from the following funding agencies and grants.
\textbf{Brazil}: Conselho Nacional de Desenvolvimento Cienti\'ifico e Tecnol\'ogico (CNPq); Funda\c{c}ão de Amparo \`a Pesquisa do Estado de Rio de Janeiro (FAPERJ); Coordena\c{c}ão Aperfei\c{c}oamento de Pessoal de N\'ivel Superior (CAPES).
\textbf{China}: NAOC, National SKA Program of China (Grant No.~2020SKA0110200); National SKA Program of China (Grant No. 2025SKA0110100); Project for Young Scientists in Basic Research of Chinese Academy of Sciences (No.~YSBR-061); Program for Innovative Talents and Entrepreneurs in Jiangsu, and High-end Foreign Expert Introduction Program in China (No.~G2023061006L); China Scholarship Council (No.~202306010363).
\textbf{Denmark}: Villum Fonden (project no.~29388).
\textbf{France}: ``Emergences'' Programme of Sorbonne Universit\'e; France-China Particle Physics Laboratory; Programme National des Hautes Energies of INSU; for IAP---Agence Nationale de la Recherche (``APACHE'' ANR-16-CE31-0001, ANR-23-CPJ1-0103-01), CNRS Programme IEA Argentine (``ASTRONU'', 303475), CNRS Programme Blanc MITI (``GRAND'' 2023.1 268448), CNRS Programme AMORCE (``GRAND'' 258540); Fulbright-France Programme; IAP+LPNHE---Programme National des Hautes Energies of CNRS/INSU with INP and IN2P3, co-funded by CEA and CNES; IAP+LPNHE+KIT---NuTRIG project, Agence Nationale de la Recherche (ANR-21-CE31-0025); IAP+VUB: PHC TOURNESOL programme 48705Z. 
\textbf{Germany}: NuTRIG project, Deutsche Forschungsgemeinschaft (DFG, Projektnummer 490843803); Helmholtz—OCPC Postdoc-Program.
\textbf{Poland}: Polish National Agency for Academic Exchange within Polish Returns Program no.~PPN/PPO/2020/1/00024/U/00001,174; National Science Centre Poland for NCN OPUS grant no.~2022/45/B/ST2/0288.
\textbf{USA}: U.S. National Science Foundation under Grant No.~2418730.
Computer simulations were performed using computing resources at the CC-IN2P3 Computing Centre (Lyon/Villeurbanne, France), partnership between CNRS/IN2P3 and CEA/DSM/Irfu, and computing resources supported by the Chinese Academy of Sciences.


\bibliographystyle{JHEP}
\bibliography{refs}

\end{document}